\RequirePackage{lineno}
\documentclass[pagesize=auto, version=last, chapterprefix=true]{scrbook}

\usepackage{amsmath}
\usepackage{amsfonts}

\addtokomafont{disposition}{\rmfamily\mdseries}
\addtokomafont{chapterprefix}{\large}
\setkomafont{dictumtext}{\normalfont\normalcolor\itshape\setlength{\parindent}{1em}\noindent}

\usepackage{cite}
\usepackage{cancel}
\usepackage[top=1in, bottom=1in, left=1in, right=1in]{geometry}
	\usepackage[colorinlistoftodos]{todonotes} % allows margin comments
	\usepackage{soul} % allows for highlighting

\usepackage{graphicx}
\usepackage{amssymb}
\usepackage{amsmath} 
\usepackage{epigraph}
\usepackage{graphicx,float,wrapfig} 	% for embedding graphics
\usepackage{color}					% Allows picking of custom RGB colors
\usepackage[font=small,labelfont=bf,textfont=sf,
labelsep=quad]{caption}				% points hyperref to top of figure, not caption

\usepackage{gensymb}
\usepackage{caption}
\usepackage{subfig}

\usepackage{xr-hyper}
\usepackage[colorlinks,linktocpage=true]{hyperref}		% for hotlinking all \cite{} and \ref{}

\makeatletter
\newcommand*{\addFileDependency}[1]{% argument=file name and extension
  \typeout{(#1)}
  \@addtofilelist{#1}
  \IfFileExists{#1}{}{\typeout{No file #1.}}
}
\makeatother

\hypersetup{						% Hyperlinked reference setup; can use RGB color values
  citecolor= black,
  linkcolor=blue,
}

\usepackage{setspace}   

\usepackage [english]{babel}
\usepackage [english = american]{csquotes}
\MakeOuterQuote{"}

\doublespace                   				%   change line spacing

\usepackage[protrusion=false,expansion=false]{microtype}

\providecommand{\keywords}[1]{\textbf{Keywords: } #1}

\title{
Mobile defects born from an energy cascade shape the locomotive behavior of a headless animal \\ \Large {\textsl{Part 3 (of 3): Organism scale}}
}

\author{
Matthew S. Bull$^1$, Manu Prakash$^2$\\ 
\normalsize $^1$ Department of Applied Physics \\
\normalsize $^2$ Department of Bioengineering, \\ \\
\normalsize Stanford University, Stanford, CA \\\\
\normalsize *To whom correspondence should be addressed: \\ 
E-mail: manup@stanford.edu
}
\date{\today}

\begin{document}
\maketitle
\begin{quote}
\footnotesize

%\begin{abstract}
%[Merge first three into simpler sentence] Organismal biophysics endeavors to link subcellular processes with their ecological implications through mechanistic understanding of organism-scale dynamics. With recent progress, the field is poised at a crossroads between low-dimensional dynamics and highly detailed computational models. There is hope for unification of these approaches with further progress bridging the details of complicated construction with the identification of hidden emergent simplicity.
The physics of behavior seeks simple descriptions of animal behavior. The field has advanced rapidly by using techniques in low dimensional dynamics distilled from computer vision. Yet, we still do not generally understand the rules which shape these emergent behavioral manifolds in the face of complicated neuro-construction --- even in the simplest of animals. In this work, we introduce a non-neuromuscular model system which is complex enough to teach us something new but also simple enough for us to understand. In this simple animal, the manifolds underlying the governing dynamics are shaped and stabilized by a physical mechanism: an active-elastic, inverse-energy cascade. Building upon pioneering work in the field, we explore the formulation of the governing dynamics of a polarized active elastic sheet in terms of the normal modes of an elastic structure decorated by a polarized activity at every node. By incorporating a torque mediated coupling physics, we show that the power is pumped from the shortest length scale up to longer length scale modes via a combination of direct mode coupling and preferential dissipation of higher frequency modes. We use this result to motivate the study of organismal locomotion as an emergent simplicity governing organism-scale behavior. To master the low dimensional dynamics on this manifold, we present a zero-transients limit study of the dynamics of +1 or vortex-like defects in the ciliary field (which is experimentally supported for small organisms). We show, experimentally, numerically and analytically that these defects arise from this energy cascade to generate long-lived, stable modes of locomotive behavior. Using a geometric model, we show how the defect undergoes unbinding. We extend this framework as a tool for studying larger organisms with non-circular shape and introduce local activity modulation for defect steering. We expect this work to inform the foundations of organismal control of distributed actuation without muscles or neurons.
\\

\textsl{Significance:}\\

Organismal biophysics is poised at a crossroads between low-dimensional dynamics and highly detailed computational models. Our work contributes to this conversation by demonstrating an 'emergent simplicity' embodied by low-dimensional behavioral manifolds in an animal without muscles or neurons. We use both top-down approaches (unsupervised inference) and bottom-up (models by construction) to show that these manifolds are stabilized by an active-elastic, inverse-energy cascade. We develop a model for the steady state dynamics of a mobile defect, which is a direct consequence of this physical mechanism. This model illustrates how the defect can be trapped and manipulated in an active elastic sheet. Finally, we show that the low variance modes of our unsupervised construction have critical consequences on the decision making of an animal without a brain. We expect that this work will be of interest to physical ethologists and the makers of intelligent machines.
%\end{abstract}

% Insert keywords here
\setlength\parindent{.45in} \keywords{Active matter, physical ethology, low dimensional dynamics, coherent structures}
\end{quote} 

\begin{quote}
\textbf{Author Contributions}\newline
M.S.B. and M.P. designed research; M.S.B and M.P. collected the data; M.S.B. analyzed the data; M.S.B. developed theory and simulations; M.S.B. wrote and M.P. edited the paper.

\textbf{Note to readers}\newline
This work is the second of three complementary stories we have posted together describing multiple scales of ciliary flocking. While we encourage readers to read all three for a more complete picture, each of these works are written to be as self-contained as possible. In the rare case where we evoke a result from another manuscript, we make effort to motivate it as a falsifiable assumption on the grounds of existing literature and direct readers to the relevant manuscript for a more in-depth understanding.

\textbf{Part 1}\newline \textsl{Excitable mechanics embodied in a walking cilium} (posted on arXiv concurrently)\cite{bullpart1} describes the emergent mechanics of ciliary walking as a transition between ciliary swimming and ciliary stalling with increasing adhesion. 

\textbf{Part 2}\newline \textsl{Non-neuromuscular collective agility by harnessing instability in ciliary flocking} (posted on arXiv concurrently)\cite{bullpart2} reports the implications of an effective rotational degree-of-freedom governing the direction of ciliary walking for agile locomotive behavior in an animal without a brain. 

\textbf{Part 3}\newline \textsl{Mobile defects born from an energy cascade shape the locomotive behavior of a headless animal} (this work) describes the emergent locomotive behavior of the animal in terms of a low dimensional manifold which is identified by both top-down and bottom-up approaches to find agreement.
\end{quote}
%=======================================================================
%\linenumbers\modulolinenumbers[1]
%\resetlinenumber[1]

%========================================================================
\bibliographystyle{naturemag}
%\bibliography{bibfile}
%=======================================================================

%\input{Part3_InverseEnergyCascade}

%\tableofcontents

\section{Introduction}

Parsing the complexity of animal behavior promises to teach us important lessons about how the living form copes with a sometimes unpredictable environment. With the development of a sundry of modern techniques \cite{berman_predictability_2016, ahamed_capturing_2021, Gilpin2017, stephens_dimensionality_2008, brown_ethology_2018, daniels_automated_2019, bongard_automated_2007, sponberg_dual_2015}, the challenge of unbiased behavioral quantification has begun to suggest a deeper underlying structure which may be exploited to further our understanding of the physics of behavior \cite{berman_predictability_2016, brown_ethology_2018, sponberg_emergent_2017}. Perhaps the most tantalizing outcome of this collective enterprise, is the growing support for paradigm of emergent simplicity \cite{transtrum_perspective_2015}. Oftentimes, from a complicated construction of a developmentally encoded yet highly individualized ensemble of many sensors, neurons, and muscles, the dynamics of the animal can be captured in an effectively much lower dimensional space of behaviors \cite{berman_predictability_2016, ahamed_capturing_2021}. There is growing support for the conjecture that the shape of these emergent spaces is enforced by evolutionary outcomes which lead to the robust self-organization of these attracting manifolds \cite{ratliff_neuronal_2021, nunns_signaling_2018, sponberg_emergent_2017}. These effective dynamics may comprise a new space in which to consider the role of the environment and the manifestation of a particular state on the output behavior. Here, we adopt a simple animal of study to develop a complementary perspective which is sufficiently distinct from the view given by the exciting and informative work in worms\cite{stephens_dimensionality_2008, ahamed_capturing_2021}, flies\cite{berman_predictability_2016} and fish\cite{orger_zebrafish_2017}. Similar approaches have been applied to collections of organisms including birds\cite{attanasi_information_2014}, fish\cite{tunstrom_collective_2013} and collections of cells \cite{copenhagen_frustration-induced_2018} focusing largely on the modes of the collective behavior ranging from flocking, milling and swarming.  

This research program focused on distilling critical lessons from simple animals has value beyond the domain of these simple nervous systems. This enterprise builds useful conceptual tools and technology to transform hard drives of data into effective degrees of freedom with hints toward more general principles. One such connection might arise in the bridge between dynamics in neural sub-spaces \cite{vyas_computation_2020} and their behavioral manifestations. The discovery of such motifs may inspire new ways toward understanding data in higher brains. With each new publication finding low dynamics, a new question emerges: is the emergence of tractable simplicity from complicated construction a theme to be exploited? \cite{mora_are_2011, sponberg_emergent_2017, brown_ethology_2018}

Yet, distributed actuation and decision making -- the hallmark of multicellular life -- is seemingly at odds with these low dimensional outcomes. This disconnect between cellular degrees of freedom and organismal behavior comes to a point in the tension between emergent low-dimensional models and models built from the ground up where the details seem to matter \cite{wang_neuron_2020}. By studying an organism without a neuromuscular system, we build our intuition slowly toward more general principles mapping the collective behavior of complicated construction onto an organism-scale emergent simplicity. In doing so, we complement existing and ongoing work on more traditional model organisms and hope to lend support to a tantalizing link between emergent animal behavior\cite{brown_ethology_2018}, a growing field in cellular cognition \cite{wan_origins_2021, brette_integrative_2021, jekely_chemical_2021} and the physics of active matter\cite{marchetti_hydrodynamics_2013,shankar_topological_2020, alert_active_2021, tu_sound_1998, cavagna_silent_2015}. 

In this work, we study the spatio-temporal dynamics of a simple animal without muscles or neurons, \textit{Trichoplax Adhaerens}, from top-down and bottom-up. Meeting in the middle, we show how the emergent behavioral manifolds of this simple animal are stabilized by the inverse energy cascade pumping energy into increasingly collective modes. By looking past the shape of the manifold to the underlying physical constraints shaping it, we demonstrate a simple system where the emergent locomotive behavior can be cast in the form of quasi-particle-like, defect dynamics in the zero-transients limit. We show that for a wide range of organism size this formulation holds and it gives us a perspective on how low dimensional dynamics can arise from a collection of individual cells and how local cellular control can influence organism scale locomotive dynamics.

\subsection*{Review on defect dynamics and inverse energy cascade}

The active matter community has long recognized the value of studying the properties of emergent quasi-particles in the form of defects\cite{shankar_topological_2020}. Defects are a convenient tool for parameterizing continuous fields and this pedigree is undoubtedly shaped by the fields' close connections to hydrodynamics and more generally, condensed matter physics in which theories like the BKT phase transition mediated by vortex gases carry fundamental importance\cite{kosterlitz_ordering_1973}. This deep relationship with patterns of interacting defects has also linked it to the ideas of turbulence in which the momentum terms are conceptually replaced with activity at low Reynolds number giving rise to spatio-temporally rich flows reminiscent of the momentum rich solutions of Navier-Stokes\cite{dunkel_fluid_2013, alert_active_2021, blanch-mercader_hydrodynamic_2017}. 

Defects are not purely mathematical tools and have important biophysical implications such as controlling cell death and extrusion from a tissue monolayer\cite{saw_topological_2017}, localizing proliferation\cite{guillamat_integer_2020}, and guiding collective density \cite{kawaguchi_topological_2017}. Thus the control of defects represents an attempt at harnessing the power of active matter to do work at the collective scale \cite{zhang_spatiotemporal_2021, peng_command_2016, gong_engineering_2020, dendresen_topological_2021}. As we develop these technologies further, we can benefit from insights into new ways to shape and control these defects in a polarized active matter. For this, we can learn much from \textit{T. Adhaerens}.

\section{Experimental signatures of defect dynamics}
\textit{T. adhaerens} is an emerging model in tissue mechanics both for its highly visible and dynamics actomyosin contractions \cite{armon2018ultrafast, armon_epithelial_2020} and its relative simplicity in components\cite{Grell1971TheSchulze, Smith2014}. The abundance of cilia also makes the organism a compelling system for understanding the coordination of ciliary dynamics in non-neuromuscular locomotion\cite{Smith2015CoordinatedSynapses., smith_coherent_2019, Senatore2017NeuropeptidergicSynapses., Grell1974ElektronenmikroskopischePlacozoa}  The spatio-temporal dynamics of the tissue displacement field is a prime opportunity to not only identify any signatures of low dimensional representations, but also begin to understand what shapes those manifold sub spaces. 

To capture the locomotive dynamics of this organism in this work, we leveraged two primary classes of experiments. In the first, we honed in on single organelle resolved imaging of the bottom of the organism to watch the spatial evolution of the ciliary dynamics. In other work, we report that these dynamics can be understood as sub-second ciliary reorientations [part 2] of cilia which are walking on the substrate [part 1]. These collective dynamics result in striking defects in the ciliary field which diffuse around at high speeds.

In the second set of experiments, we stain the tissue with a lipid loving florescent dye which absorbs preferentially into a sparsely sampled cell type comprising about 10\% of the bottom tissue\cite{Smith2014} using techniques described by Prakash et al \cite{prakash_motility-induced_2021}. These uniformly spaced fluorescent spots make an idea platform on which to analyze the flow field using particle image velocimetry \cite{Thielicke2014PIVlabMATLAB}. This gives us a direct measurement of the time evolution of the local tissue displacement field over the entire organism. 

When we plot a local measure of the field polarization we find evidence for highly localized domains with high disorder reminiscent of point defects. These point defects are identifiable by a couple of key signatures. The first is that the defects sit at the center of a rotational coherent structure which is highly vortex like. This means that when we integrate the phase around a path in this field we enclose one quanta of topological charge. We use this definition to call the defect a +1 defect within the orientation field. Because in the limit of small organisms, this locomotion becomes approximately solid body (e.g. the activity scale is not strong enough to deform appreciably the weakest mode of the tissue [part2]), the rotation deviates from the classic localized defects in fluid mechanics, where the shear exceeds the viscosity at a critical radius and sets a finite size vortical flow with a 1/r fall-off outside of this critical radius \cite{acheson_elementary_1990}. Instead in our polarized active solid, the effective radius of the rotation speed grows with time until it achieves full system size.

Another key observation is that even when the defects move at high speeds, they remain quite localized only picking up small strands of disorder in the direction orthogonal to the defect travel direction. This is an important characteristic informing the low dimensional dynamics of the organism, as we can imagine many alternative ways to distributing the local orientation such that the net polarization is zero. This supports the notion that we will make precise in the coming sections that the self-organizing dynamics push active forcing injecting on the shortest lengthscale up to collective modes with the longest possible lengthscale in the problem through a combination of preferential dissipation and energy passing between modes [part 2].

A final broad stroke observation we report is that these defects are effectively bound to the organism and in the absence of strong activity or environmental noise have a tendency to self-center if near enough to the center. This trapping of the defect by relative to the organisms' boundary suggests that (at least in sufficiently small animals) the defect is strongly influenced. 

\begin{figure}
\begin{center}
\includegraphics[width=0.6\textwidth]{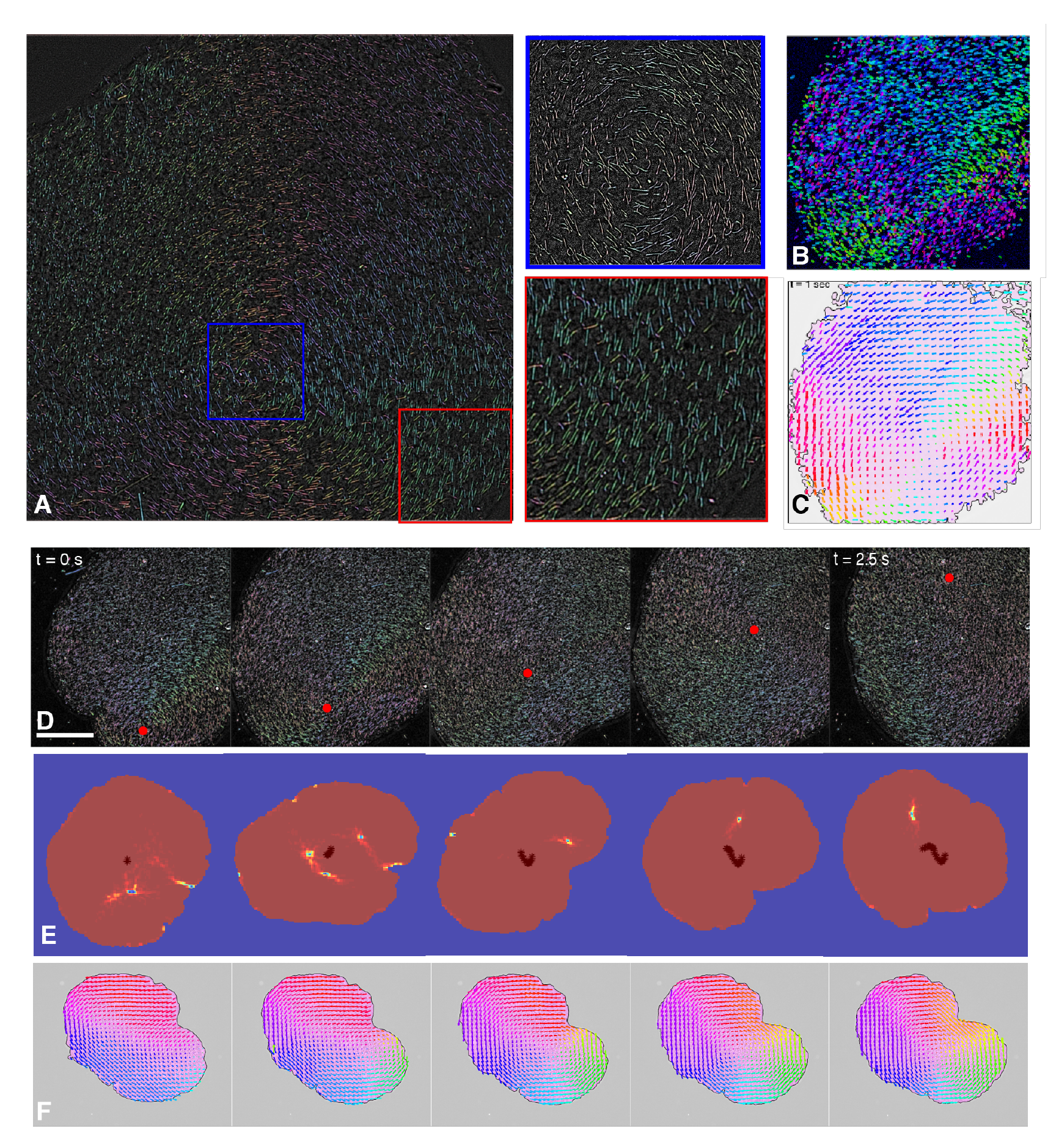} 
\caption{\label{fig:figure_1_experimentalvortex} Ciliary flocking underlies the collective locomotion of \emph{Trichoplax adhaerens} from hundreds to millions of cells (A) Wide field yet single cilia resolved imaging of entire organisms (60x Bright Field optical sectioning at 100 fps) reveals collective ciliary flocking with defect structures. These defects are characterized by regions of high orientational disorder in the ciliary field. The disorder reduces as the distance from the center of the defect grows. %Second, around these defects, we observe strong counter-clockwise rotation of the cell bodies with alignment of the ciliary field.
(B) We visualize the entire organism's ciliary orientation field, $\phi$, using a symmetry based ciliary segmentation (see methods, color scheme red-blue-green-red color scheme from -$\pi/2$ to $\pi/2$). 
(C) From the same movie, we simultaneously extract the motion of the cell bodies using a digital image correlation measurement on the low pass filtered image (see method). This displacement field is colored by their vector orientation using a colormap from -$\pi$ to $\pi$ from Red-blue-green-red-blue-green-red to match the color to angle of the ciliary orientation field above.
(D) Ciliary orientation field of a walking animal show rich dynamics which revolve around the position of the defect position within the ciliary field (defect colored with a red dot to guide the eye). In this sequence, we observe the defect rapidly traveling from the bottom of the organism to the top in a timespan of 2.5 seconds. Following the removal of this defect, the organism's ciliary field converges to a symmetry broken state. This transitions an animal from a rotational state to translation state in behavior space. 
(E) Complementary techniques allowing us to track a sparse cell type in the bottom epithelium, permit the reconstruction of the sparsely labeled cellular displacement field $\theta$. This cell displacement field can be colored using the same colormap spanning $-\pi$ to $\pi$ from Red-blue-green-red.  
(F) To observe defect dynamics impact on the cell displacment field, we translate the vector field into a local order map by calculating the local polarization: $P_{i, local} = \sum_{j \in \text{nearest neighbors}} cos(\theta_i - \theta_j) / N_{\text{nearest neighbors}}$. In the local-polarization field the defects show up as yellow points in a red background. This visualization shows that these defects are highly motile exhibiting generation, dissipation and merging. Intriguingly, many of these defects are self centering for sufficiently small organisms up until they reach a critical distance away from the center in which case they are rapidly expelled. Other sources of noise in the cell displacement field can result in defect pairs which quickly annihilate. }
\end{center}
\end{figure}
\newpage

\section{A simple physical model gives a touchstone against which we can compare the dynamics of the animal}

The goal of this section is to motivate the simplest possible many-cell model with close models reported throughout the literature \cite{ferrante_elasticity-based_2013, szabo_phase_2006, copenhagen_frustration-induced_2018} as a baseline. The value of comparing the experimental dynamics to these collective mechanics builds our intuition for what in the observed mechanics should be surprising and what should not. This will enable future studies where we can focus on the differences between the model and the organismal behavior to look for signatures of chemical control and adaptation. 

This model comprises of a set of degrees of freedom in 2D space representing a collection of cells, $|r\rangle$ in a highly damped medium. We define a fixed topology network of springs spatially connected into a lattice where each cell has 6 neighbors. On each one of the nodes of this network, we have an active force vector which has both a heading and an amplitude encoded by the vector $|\phi\rangle$ for each cell -- this vector represents the simplest manifestation of the cell's internal state. Based upon parallel work [part 2], we posit that this cellular state is updated through the applied torque exerted on the cell by the rest of the tissue pulling it into line (akin to a castor wheel). Thus the dynamics of the tissue at each point is determined by the sum of the active contributions with the elastic contributions:
\[
\partial_t |r\rangle = |\phi\rangle - \nabla_r E(|r\rangle)
\]
while the activity vector updates through two equations:
\[
\partial_t | \hat \phi \rangle = -\Gamma | \hat \phi \rangle \times  \nabla_r E(|r\rangle)
\]
\[
\partial ||\phi|\rangle = -\tau (||\phi|\rangle - a_o) + |\xi\rangle
\]
The activity amplitude in this case behaves like a Orenstien-Uhlenbeck noise process driven by an activity noise centered around a preferred activity. 

The bulk dynamics can be characterized by watching the evolution of the activity orientation field which is governed by fast timescale reorientation statistics and the amplitude field which is driven stochasticially. The third field which we visualize is a local force on the tissue field which emerges from the gradient of the elastic free energy of the tissue in response to the active forces. We propose thinking of this local tissue force as a measure of the integrate disagreement between the activity vectors of nearby cells which highlights long-lived defects in the underlying activity field. These dynamics can be plotted as:
\begin{centering}
\begin{figure}[h!]
    \includegraphics[width = \textwidth]{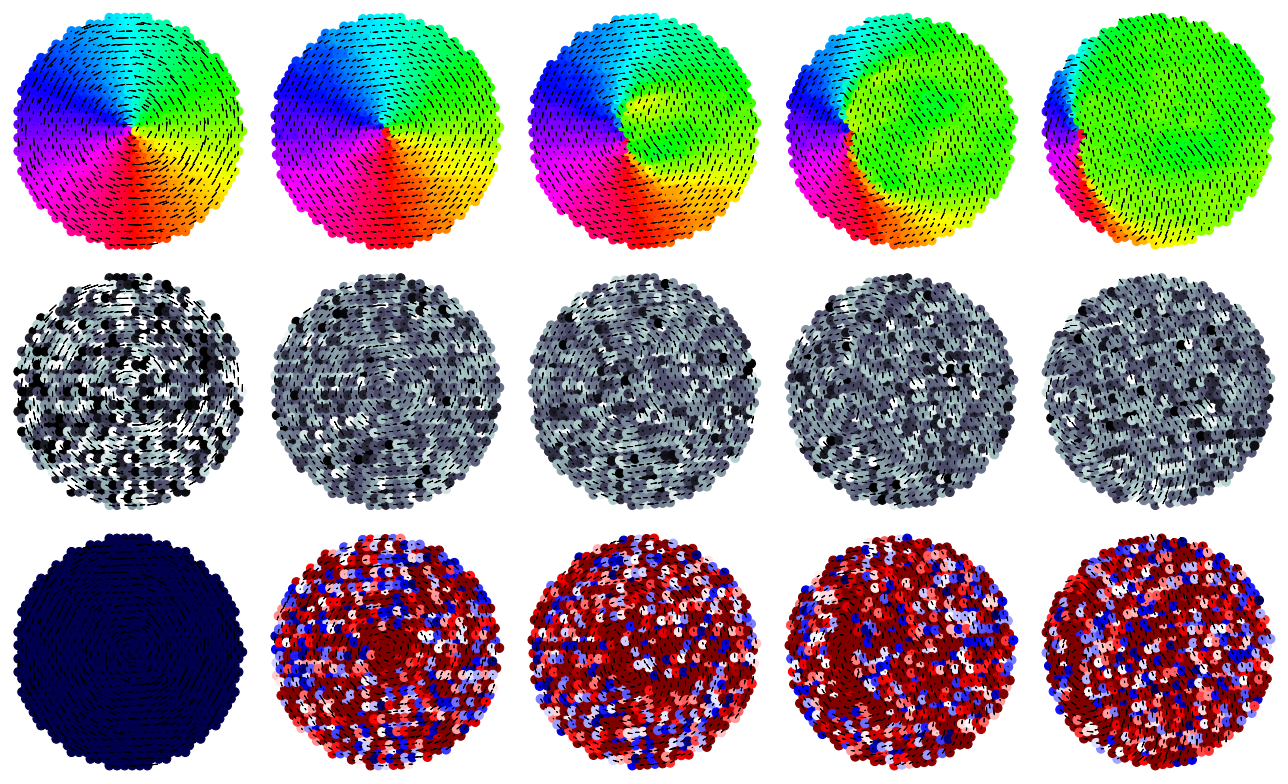}
    \caption{The time evolution of the spatio-temporal dynamics show the motion of a mobile vortex in three simultaneous channels: the activity orientation field, the activity magnitude field, and the locally evaluated tissue force.}
\end{figure}
\end{centering}

This minimal model exhibits a direct analogy to the classic flocking transition with increasing activity noise amplitude playing the role of an effective temperature. A a critical effective amplitude of activity fluctuations, the  tissue's polar order drops toward zero as the noise overcomes any ability to order the polarization of the dynamics. This result is complementary to the results in the literature where a similar qualitative phenomena is show for 'orientational noise' \cite{szabo_phase_2006} which acts on the heading directions of the activity vectors and 'positional noise' \cite{ferrante_elasticity-based_2013} which acts directly on the cells themselves. In the limit, where activity is large compared to environmental coupling induced fluctuations, the source of the noise generates qualitatively similar phases in the simplest possible model.

\begin{centering}
\begin{figure}
    A)\includegraphics[width = \textwidth]{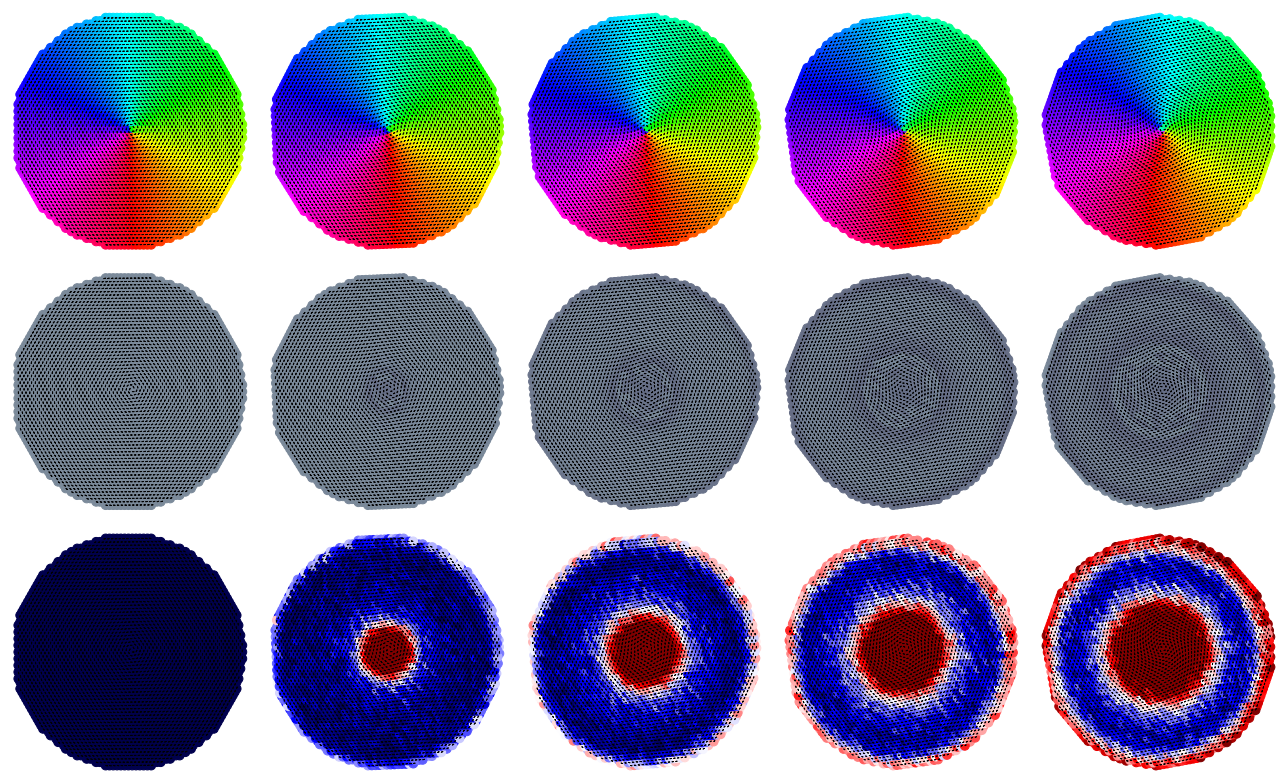}
    B)\includegraphics[width = 0.3\textwidth]{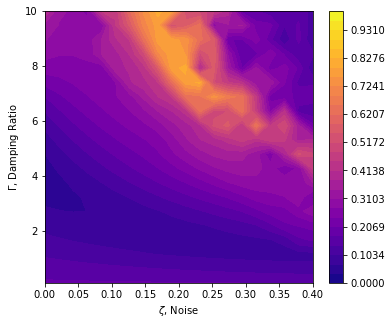}
    C)\includegraphics[width = 0.3\textwidth]{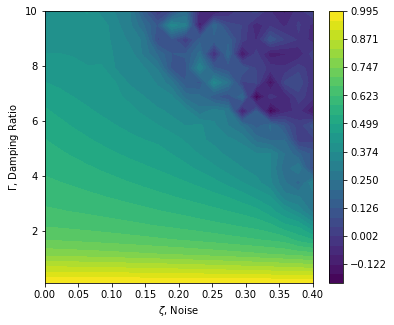}
    D)\includegraphics[width = 0.3\textwidth]{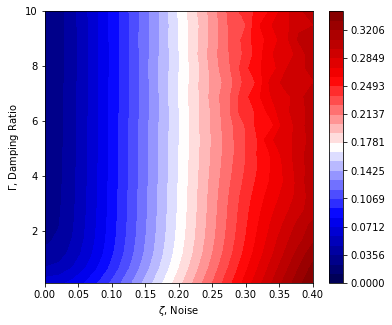}
    \caption{A simple phase space of the noisy polarized active elastic sheet shows three qualitative regimes: i) long-lived and stable vortex, ii) a polarized locomotion regime, and iii) a disordered regime. A) By visualizing the orientation, activity amplitude (low noise limit) and the locally evaluated tissue force through the first 1000 steps (equally spaced) of simulation on $10^4$ cells, we see how the force in the tissue is forced to compensate for the disagreement between the activity and the rigid body rotation of the tissue in a vortex state.  B) Summarizing the mean polarization at long times across 1e3 simulations illustrates these three regimes. At low noise and low ratio of timescales, $\Gamma$, the vortex state is very stable for long times with self-centering behavior resulting in low polarization. At low noise and higher $\Gamma$, the vortex core is more susceptible to diffusion and can sometimes escape its bound state at the center of the tissue. As noise increases the mean lifetime of the vortex state decreases and the escape can happen at lower and lower values of $\Gamma$. Above a critical noise, mean polarization drops indicating a transition into a disordered regime of the phase space. C) The vortical polarization of the tissue displacement, $L = \sum_i \hat a_i \times \vec r_i$, confirms this story with high vortical polarization at low $\Gamma$ and low noise, which crosses over to low measured vortical polarization at high noise and high $\Gamma$. D) The phase space for the tissue force shows that for $\Gamma > 1$, the force in the tissue -- a measure of integrated disagreement -- grows in proportion with the activity noise, $\langle F \rangle \propto \zeta$}
    \label{fig:activitynoisephase}
\end{figure}
\end{centering}

With this minimal model in place, we can go beyond these summary statistics to study the resulting spatio-temporal dynamics of the tissue. We present this work with the aim of presenting the highly non-trivial dynamics of an active elastic sheet so that we can better appreciate where to be surprised when observing the collective dynamics of the living \textit{T. Adhaerens}. 

\section{Describing the spatio-temporal dynamics}
Placozoa and inspired models of polarized active elastic sheets exhibit rich spatio-temporal cell displacement dynamics. These collective responses walk the boundary between being stable enough to maintain cohesion yet sensitive enough to illicit large system wide responses to local environmental stimulus. These collective dynamics leave much to study in their own right as the temporal progression between cell displacements is governed by a highly dynamic system in the limit of rapid reorientation of the activity vector seen in placozoa\cite{bullpart2}. 

To begin to get a handle on this spatio-temporal complexity, we take an unsupervised decomposition approach inspired by Gilpin \textit{et al}\cite{gilpin2017vortex} which permits us to study the dynamics of these polarized active elastic sheets. We begin by collecting long, microstate aware simulations of the polarized active sheet with fast reorientation times, $\Gamma/\gamma \sim 15$. The \textit{in silico} organism gives a record of the cell displacement vector and position every agent in the tissue, $\theta_i(t)$, and $r_i(t)$. From these records, the net displacement of the organism is computed at every time instance defining the organism's heading direction and speed, $\langle \theta \rangle (t)$. We use this direction to rotate the real space representation of the cell-positions and displacements such that the heading is held constant at 0 radians, $\tilde \theta_i(t) \rightarrow \mathbb{R}(\langle \theta \rangle (t))\theta_i(t)$ and $\tilde r_i(t) \rightarrow \mathbb{R}(\langle \theta \rangle (t)) r_i(t)$, where $\mathbb{R}(angle)$ is the rotation operator. This approach mimics the head registration done by previous authors decomposing the behavioral repertoire of worms\cite{stephens_dimensionality_2008}, and flies\cite{berman_predictability_2016} and can be thought of as the amplitude of the modes around the spontaneously symmetry broken state since this tissue does not have the same kind of developmentally broken symmetry between head and tail.

Next, we define a grid in the organisms heading frame of reference. This grid underlies a uniformly spaced displacement field which we use to define a displacement pattern at any single point in time. We construct the values on this grid by taking the mean cell displacement vector of the set of cells, $r_i(t)$, which map to each grid point $g_i$ as its nearest out-of-set neighbor giving $\Theta_i$ at every grid point at each timestep. The utility of this structure is that is allows us to simply encode and decode the spatial component of our pattern into the identify of a fixed in place structure. By making a map between identity and location for encoding, we can account for the spatial structure more simply in the unsupervised operation. This allows us to treat the dynamics at the evolution of a sequence of numbers in time opening up to more simple types of analysis including Dynamic mode decomposition and variants of Principle Component analysis.

\begin{figure}
\begin{center}
\includegraphics[width=0.75\textwidth]{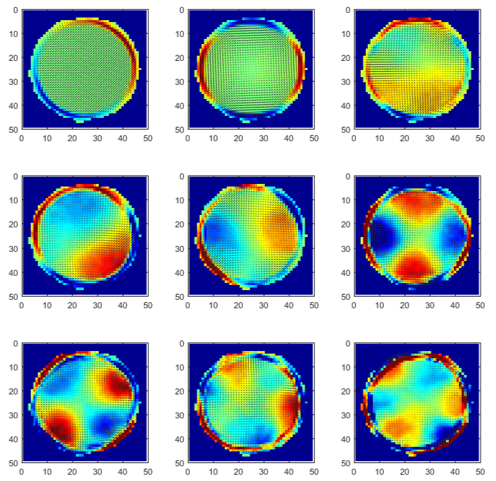} 
\caption{The first 9 modes reconstructed from a Principle-Component-Analysis-inspired method show the dominant cell-displacement fields over long observations of \textit{in silico} trajectories of small tissues, $n \sim 1000$. The color represents the local curl of the displacement field highlighting canonical structures from the spontaneous symmetry broken state to the rotation state. Each mode is ordered by its eigenvalue which is related to the percent of observed variation.}
\label{fig:9modes}
\end{center}
\end{figure}

To turn the sequence of grid points vectors into numbers, we use the mapping between the dot product in two dimensions and the product of a complex number with its complex conjugate. This allows us to write each rotated displacement vector as: $z_i = \Theta_i\cdot\hat x + i\Theta_i\cdot\hat y$. We then write each time-point in the observed sequence as a new column in the data matrix for a total of $T$ observations. If there are $N$ grid points sampled over $T$ observations the data matrix is of the form of an $N \times T$ matrix, $\mathbb{D}$, with each element $\in \mathbb{C}$. By subtracting off the mean (giving $\tilde{\mathbb{D}}$), we can represent the covariance matrix as the normalized conjugate transpose product, $\tilde{\mathbb{D}}^{\dagger} \tilde{\mathbb{D}}$. Solving for the eigenvectors and eigenvalues of this matrix gives us the principle components and their directions. We can then apply the encoding operation in reverse to take the eigenvectors from complex numbers to vectors and then remap them into space through the identification of grid points. When this is completed on data generated from a long polarized active elastic sheet trajectory $10^7$ timesteps for $10^3$ cells initialized in a hexagonal lattice, we reconstruct the first 9 modes observed in figure \ref{fig:9modes}.

Complementary to the mode shapes, we also gain access to how much each mode represents the observed dynamics through direct calculation of the eigenvalues. Plotting up the distribution of mode powers represented, we find that the first two modes (a translation mode and a rotation mode) dominate the dynamics strongly. After the first two modes, a log-log plot of the mode power versus rank reveals a two-regime powerlaw like relationship.

\begin{figure}
\begin{center}
\includegraphics[width=0.40\textwidth]{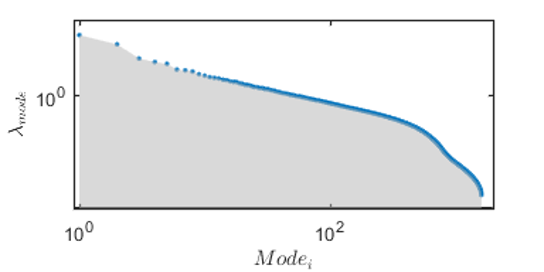} 
\includegraphics[width=0.55\textwidth]{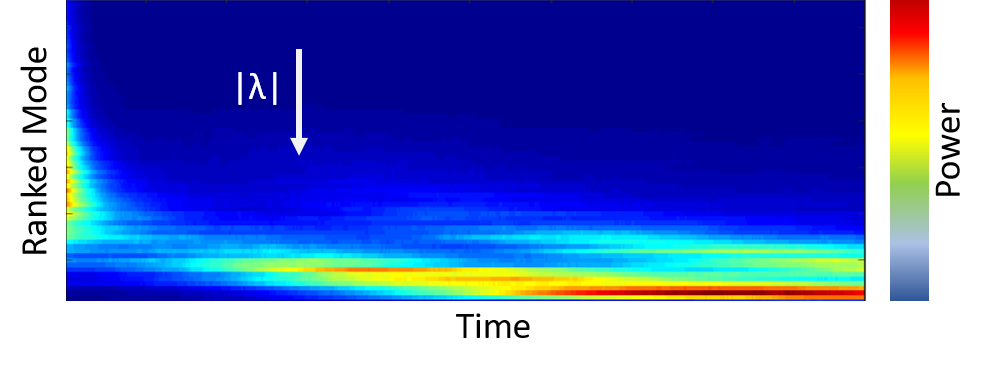}
\caption{The statistics of representation power is represented by the eigenvalues. A) Plotting the mode eigenvalue versus the rank reveals a two-regime powerlaw like relationship. Notice that the first two modes, the translation and the rotation capture the bulk of the representation. Following these two dominant modes, we find a first Zipf-like powerlaw in the rapidly decaying eigenvalues. B) Using the unit-vectors, we can capture the amplitude of each of these modes in the time-course by leveraging the inner-product of each mode's eigenvector with the observed displacement. Upon random initialization of the active orientations, the modes are broadly active. However, as time progresses power is pumped from modes with low representation power toward the modes with the highest. This is directly reminiscent to the inverse energy cascades seen in randomly diffusing active vectors in tissues\cite{henkes_dense_2020} and active crystals \cite{ferrante_elasticity-based_2013}}. 
\label{fig:Power}
\end{center}
\end{figure}

The first powerlaw decays slowly across the first 10's of modes and then more rapidly to the end. The initial Zippian-like distribution is reminiscent of expressive complexity and suggests that even short-lived modes have important expressive power in the spatio-temporal dynamics a polarized active elastic sheet with activity noise. 

Adopting the alternative perspective, it is encouraging that a large proportion of the observed dynamics can be captured effectively with relatively few dimensions. For the studied small organisms with $10^3$ cells (\textit{in silico} and live), the first two modes capture 60\% of the variation. In the following work, we develop two simplified frameworks to understand the locomotive implications of these longest wavelength modes as a reduced order understanding of the implications of these collective modes on locomotive transport. 

Figure \ref{fig:Power}B shows the dominant importance of the most representative modes which effectively captures the vast majority of the long time behavior of the tissue. Most of the energy being injected at the cellular (shortest lengthscale of the problem) lengthscale cascades down to these few dominant modes consistent with the two mechanisms of direct mode coupling and preferential dissipation of higher stiffness and shorter lengthscale modes. 

This observation motivates the following work where we develop two simplified frameworks to understand the locomotive implications of these longest wavelength modes as a reduced order understanding of the implications of these collective modes on locomotive transport. These reduced order models have important implications for the locomotive dynamics of the organism, which we highlight through the study exploration in a zero-information environment. 

\subsection{Emergent dynamics of the two longest lived modes}

First, we explore a low dimensional model reflective of the locomotive dynamics of the spontaneous symmetry broken state. This work demonstrates a smooth crossover between an active-elastic resonator and the excitation of a Goldstone-like mode associated with the spontaneous symmetry breaking. The energy transfer between the two modes is controlled through a dynamical link between the timescale of influence and the resonator frequency. This coupling between the modes is mediated by the distinct timescales of the problems and can be modeled via a smooth crossover between the dominate response of the tissue to a measured perturbation. 

Both experimentally and numerically, the dynamics of the next leading order pattern is governed by the evolution of a vortex-like state which is punctuated by a highly mobile +1 point defect. The localization of this topological defect opens up the alternative description of the long-time dynamics (relative to the transients which scale with system size) of the organism. We propose a simple energy minimization argument which allows us a derive an effective energy landscape for vortex centering. We promote this description to an effective degree of freedom on an energy landscape coupled to the environment by a complicated noise representing the integral across the other modes. We use this model to show the transition from bound to unbound vortex position and hint toward mechanisms to control this particle position as a function of time. These opportunities for control suggest candidate hypothesizes for how the organism harnesses the control of millions of cells in service of the collective without a central neuro-muscular system.

In part 2 of this series of papers on ciliary flocking\cite{bullpart2}, we showed how the addition of a rotational degree of freedom coupled to activity fluctuations derived in part 1 could illicit an excitable active elastic system best understood as a manifestation of a parametric amplification and a saturating non-linearity. Continuing along this thread, we propose viewing these collective dynamics via a mode decomposition of the governing dynamics. Previous work in this space has suggested that for a polarized active solid, a natural decomposition basis can be defined by the normal modes of the elastic network itself\cite{Bi2016, henkes_dense_2020}. In the previous work, the activity is uncorrelated from the displacement fields allowing for a formal analysis of the power distribution between modes. High spatial frequency modes have much higher stiffnesses and characteristic frequencies causing both less amplitude of excitation and more rapid decay of displacement back to zero mode amplitude. Adding a polarized activity to the dynamics adds the lens of explicit mode coupling between excitations. The work by Ferrante et al\cite{ferrante_elasticity-based_2013} shows a compelling decay of mode amplitudes consistent with the combination of explicit mode coupling and more rapid mode decay. We adopt this viewpoint for this work where activity is considered a small perturbation to the passive dynamics and apply it to the interpretation of the energy distribution as a combination of preferential damping and direct mode coupling.

\section{Directly visualizing the attracting manifold}
The collective dynamics of this dynamical system is amenable to another analysis which does not depend strongly upon the observation of complex trajectories. The high dimensional attracting manifold can be very easily projected down to a simple space which keeps track only of the magnetization amplitude and direction. 

To leading order, the dynamics of this collection abides by the physics of a spontaneous symmetry breaking. Commonly, the amplitude is set by the minima of the effective wine-bottle potential and fluctuations in the radial direction have an small energy cost associated with the effective stiffness of the radial mode. Complementary to this massive amplitude mode, the system sterotypically generates a mass-less mode in the azimuthal direction which is commonly called a Goldstone mode\cite{attanasi_information_2014}. These massless modes allow the system to move around this basin of the wine-bottle shaped potential in response to noise. 

For sufficiently small organisms\cite{bullpart2}, these dynamics can be directly observed in a low dimensional representation of the attracting manifold which characterizes the motion of the tissue. We call this representation, vector polarization space. 
\[
\vec P = \frac{\sum_{i = 1} ^N \left[a_i\cdot\hat x + a_i \cdot \hat y\right]}{|a| N}
\]

By projecting the 2N degrees of freedom of the tissue into the vector polarization space, we can visualize the time evolution of the collective dynamics of a polarized active elastic sheet as the in-silico animal is attracted toward the hyper-sphere attracting manifold embedding in 2N dimensions. In the simplified dynamics of the numerical model, the dynamics of the tissue quickly approaches the symmetry broken state where the tissue is traveling in a direction coherently. Previous work has shown\cite{ferrante_elasticity-based_2013} that these symmetry broken states are linearly stable in active elastic sheets with distributed polarized activity. These results agree with our in-silico experiments where we see that activity noise is sufficient to drive the slow reorientation of the tissues' direction of travel through small excitations to of the Goldstone modes associated with this collective mode. 

\begin{figure}[h!]
\centering{
\includegraphics[width = 0.95\textwidth]{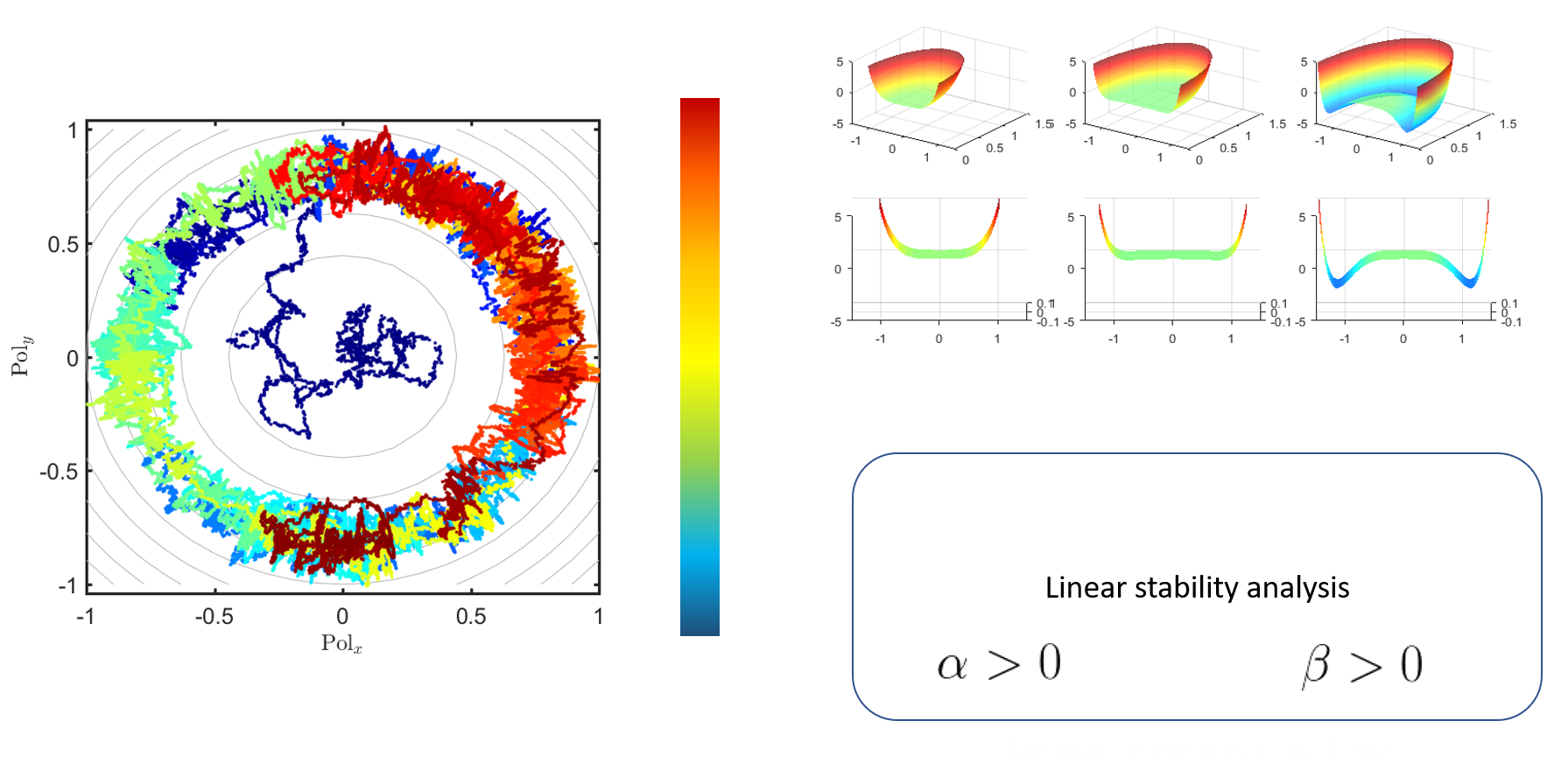}
\caption{Polarization plane}
\label{fig:polarizationplanedirection}
}
\end{figure}

In experiments, we see that for sufficiently small organisms, the dynamics are strongly attracted to the polarized hyper-sphere, but we gain new appreciation for the complex dynamics of the short timescale relaxations to this manifold. These relaxation dynamics are punctuated by highly mobile +1 defects which occupy the bulk of the low-polarization states observed in small ($ d= 500 \mu $m) animals. 

\begin{figure}[h!]
\centering{
\includegraphics[width = 0.5\textwidth]{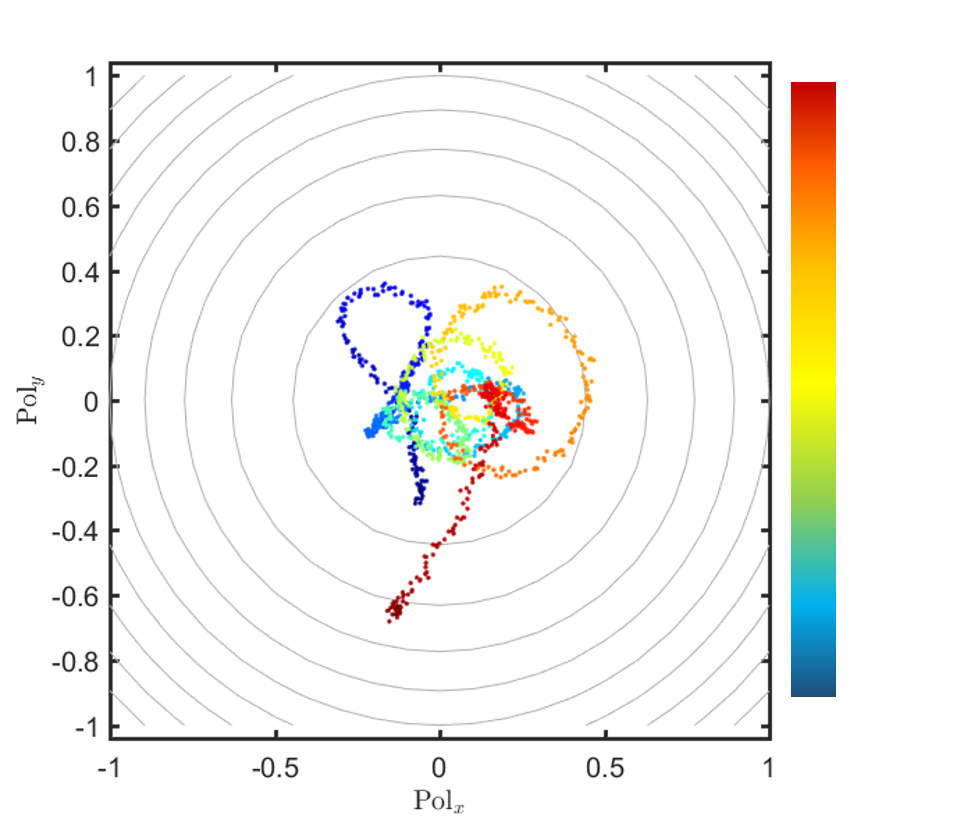}
\caption{Polarization plane from experiment}
\label{fig:polarizationplaneexperiment}
}
\end{figure}

To better classify the dynamics of these lowest lengthscale modes, we adopt heavily used metrics for the angular momentum or vortical polarization\cite{tunstrom_collective_2013, copenhagen_frustration-induced_2018}:
\[
V = \left|\frac{1}{N}\sum_i \hat a_i \times \vec \delta r_i \right|
\]
where $\hat a_i$ is the ith activity vector direction and the $\vec \delta r_i$ is the vector identifying the distance between the cell and center of the tissue. This combintation of metrics keeps track of how much the tissue is rotating and how much the tissue is translating. Rotation is distinct from the Goldstone modes by the fact that a Goldstone mode is a linear excitation of the system on the polarized hypersphere, while rotation represents a highly nonlinear excitation that excites both amplitude and aszimuthal components. 

Thus the goal of this work is to understand these highly nonlinear excitation and relaxation processes which drive the tissue toward its attractor manifold. When we plot up the dynamics in this two dimensional space characterised by $V$ and $|\vec P|$, we can begin to study projections of the relaxation trajectories. 

We plot these relaxation trajectories for a set of experiments on the simple numerical model for different choices of noise parameter and organism size. We begin this set of simulations in the +1 vortex state. As a function of the activity noise, we find three possible outcomes: i) the vortex is stable and self-centering against fluctuations (fluctuations around |P| = 0, V = 1), ii) the vortex decays along the line connecting (|P|= 0, V=1) $\rightarrow$ (|P|= 1, V=0), and iii) the vortex decays to $(|P|= 0, V=0)$. The kinetics of these relaxation processes are strongly dependent upon $\Gamma$ and in silico tissue size (N = 400 to 1e5 cells). By studying size, we find that activity noise drives larger tissues further from the relaxation manifold connecting (|P|= 0, V=1) $\rightarrow$ (|P|= 1, V=0).
\begin{centering}
\begin{figure}[h!]
    A)\includegraphics[width = 0.45\textwidth]{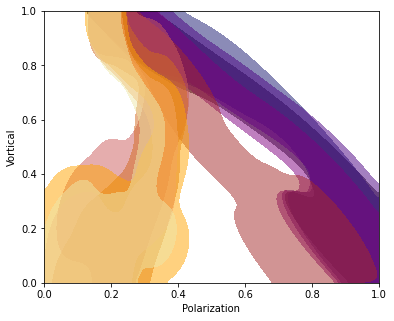}
    B)\includegraphics[width = 0.45\textwidth]{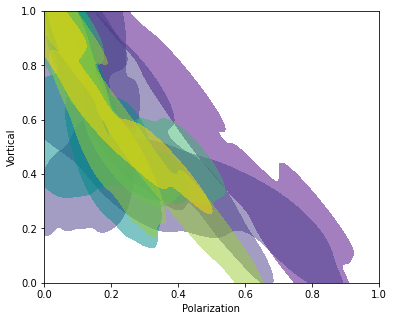}
    \caption{Numerically varying noise and size can qualitatively alter the relaxation of a vortex to polarization state. A) By varying noise from low noise (blue) to high noise (orange), we see the polarization space probability density transition from the relaxation manifold connecting $|P| = 0, V=1$ to $|P| = 1, V=0$  to the new disordered state. B) Size (1000 cells = blue, 1e5 cells = yellow) has the effect of pushing the actual trajectory further toward the disordered state than small tissue relaxations. Size seems to destabilize this manifold and reduce the magnitude of the underlying }
\end{figure}
\end{centering}

The first set of phenomena observed as we vary the noise captures the well known dynamics of the swarming, milling and flocking collective behaviors in fish\cite{tunstrom_collective_2013} and migrating cell clusters\cite{copenhagen_frustration-induced_2018}. This result shows that activity fluctuations coupled to an otherwise stable flocking state can evoke both swarming and milling depending upon the amplitude of the noise. This corroborates our earlier findings that 

The size dependence of more reminiscent of recent work where longer length-scale modes are destabilized by the interplay of cell polarity and traction force where freely growing cellular epithelia give rise to unstable modes at wavelengths $\> 1 $ mm\cite{heinrich_size-dependent_2020}.

\section{An inverse energy cascade}
The inverse energy cascade drives energy coupled at the shortest length scale toward larger and larger lengthscales. This is a powerful phenomena in forming long-wavelength collective patterns in the dynamics of 2D tissues. 

Fluid dynamics has a rich history of understanding energy transfer between scales\cite{alert_active_2021}. In turbulent flow driven at macroscopic length scales (in 3D) energy is cascade from long length-scales to short lengthscales. At sufficiently short length scales viscosity takes over and the resistance to shear flow at these short length scales is dissipated as heat into the fluid. In this way, the energy cascade shuttles energy injected at long length scales down to the Kolomorgov scale where it is converted to heat through the viscosity mediated dissipation at short lengthscales. 

Critically, nematic and polarized active fluids do not follow the same thread of 'whirls off to heat' \cite{alert_active_2021}. Commonly, active nematic and polarized active fluids are represented in strongly damped environments where momentum is rapidly (instantaneously on the timescale of the other dynamics) dissipated. The energy is injected at the shortest lengthscales of the problem which then self-organize to longer lengthscales through collaboration of the active vectors. This inverse energy cascade has the effect of decreasing the entropy production rate of the system by coupling the energy into longer-lived collective modes rather than the more rapidly dissipated high-spatial frequency modes.

At steady-state, active nematics forego energy transfer by dissipating the energy injected at each lengthscale through a combination of shear and rotational dissipation. This means that that lengthscale selection is driven by an extensional instability at the onset of active nematic turbulence\cite{alert_universal_2020, alert_active_2021, martinez-prat_selection_2019}.

In polarized active matter, we have a very different situation in which the energy driving the system out of equilibrium is injected at the shortest length scale\cite{alert_active_2021}. Collective modes are then manifestations of that energy pumping up to longer length scales via mode coupling combined with preferential dissipation of higher energy modes. We can understand a specific example as this inverse energy cascade in the context of our active elastic sheet.

The dynamics of a state of cells can be written simply as:
\[
\frac{\partial}{\partial t} \left|\theta \right \rangle = a_i | \phi \rangle + \vec \nabla_\theta E 
\]

\[
\frac{\partial}{\partial t} \left|\phi\right\rangle =  \Gamma \vec \nabla_\theta E \times |\phi\rangle
\]

Following Bi et al\cite{Bi2015a} and Henkes et al \cite{henkes_dense_2020} we can use the leading order contributions of the gradient of the free energy, to capture the small amplitude response of the system. By expanding our response around a local minimum we can represent the effective dynamics in terms of the dynamical matrix or locally evaluated hessian.

\[
\nabla E(|\psi \rangle) \approx \cancelto{0}{\nabla E|_{|\psi\rangle =|\psi_o\rangle}} \hspace{0.3 cm}+ \frac{1}{2} D (|\psi\rangle - |\psi_o\rangle)
\]

This means that when we expand around the local minimum the dynamics to leading order are dominated by the Hessian (which is this case is identical to the square of the dynamical matrix). If we apply separation of variables, the dynamics become a simple matter of eigenmode with a shared eigenvalue governing the dynamics of that mode. $D |\nu\rangle = \lambda |\nu\rangle $.

These modes define the basis of upcoming calculations:
\[
| \theta \rangle = \sum_{\nu} a_{\nu} |\nu\rangle
\]
\[
| \phi \rangle = \sum_{\nu} b_{\nu} |\nu\rangle
\]

By plugging these dynamics into our original equations gives us a new set of dynamics in terms of the mode amplitudes:
\[
\partial_t a_\nu = -\lambda_\nu a_\nu + b_\nu
\]
\[
\partial_t b_\nu = -\Gamma \sum_\mu \lambda_\mu a_\mu O_{\nu, \mu}
\]
where $O_{\nu,\mu}$ captures the orthogonal overlap between the modes. By definition the inner product of these modes is zero, and the cross product of these modes represents the coupling between the two modes. Modes which are very similar will have small amplitudes whereas modes that are very different will have strong coupling. 
\[
O_{\nu, \mu} = \sum_{i} \left[(\nu_i \cdot \hat x)(\mu_i \cdot \hat y) - (\nu_i \cdot \hat y)(\mu_i \cdot \hat x)\right]
\]

This equation shows the coupling between many amplitude modes mediated through the amplitude of the activity vector. This coupling is conceptually analogous to a symmetric synaptic coupling matrix used often in the artificial neural networks communities to represent information being passed between model neurons. In this system, the overlap matrix of the modes is shaped not by the learning dynamics of the synapse, but instead the mechanical structure of the tissue (including micro-structural disorder and the details of the local activity). 

For the purpose of this work, we present the lowest order version of this high dimensional coupled, weakly nonlinear dynamical system. One of the nice things about the active-elastic modes is under a linear approximation we can write the entire equation in terms of second order dynamical system in terms of the mode amplitude\cite{bullpart2}. This gives is the time evolution of the amplitude of the $\nu$-th mode as:
\[
\partial_t^2 a_\nu = -\lambda_\nu \partial_t a_\nu + -\Gamma \sum_\mu \lambda_\mu a_\mu O_{\nu, \mu}
\]

This is a very interesting looking dynamical system which= has the approximate form of a harmonic oscillator where the effective stiffness arises from the interaction of the many other modes. This peculiar looking equation can support instantaneous $Q$-factors for individual modes which exceed 0.5 (and thus are underdamped). These transients are the subject of other work as we showed experimentally, numerically and analytically in work in part 2\cite{bullpart2}.

In the work, we hope to study the underlying manifolds present under all these dynamics, so we will for the moment only study the steady state solutions of this equation which gives us:
\[
\sum_\mu \lambda_\mu O_{\nu, \mu} a_\mu = 0
\]
which can be viewed as a study of the null-space of the linear dynamical system. This null-space or kernel represents the number of linear combinations of the matrix which satisfies $M \vec a = 0$ where $M \equiv \Lambda O_{\nu, \mu}$ and $\Lambda$ is the matrix which results from stacking N repetitions of $\vec \lambda$. 

\begin{figure}[h!]
    \centering
    \includegraphics[width = 0.7\textwidth]{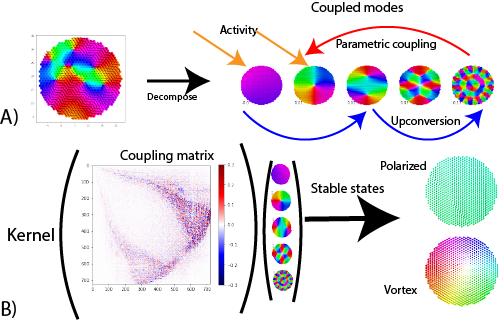}
    \caption{We identify the zero-transients stable states of the coupled active-elastic modes by studying the null-space of the emergent mode coupling matrix. A) We begin by linearly decomposing the dynamics of the active elastic sheet into a coupled active-elastic mode representation. B) Equipped with the numerically generated coupling matrix (for a crystallized adhesive granular solid), we find that the fixed points of the equation live in the numerically-constructed 3 dimensional null-space of the matrix. The first direction is the zero vector, the second and third dimensions of the kernel are the polarized and vortex states of the tissue. Using this result, we motivate the study of the attracting manifold connecting the polarized and vortex states.  }
    \label{fig:stablestatesfrominverseenergycascade}
\end{figure}

This activity mediated coupling matrix can be solved for numerically for the case of a triangular grid of a discrete elastic solid with monodisperse particles following the methods of Prakash et al\cite{prakash_motility-induced_2021}. The resulting normal modes of the elastic structure can be used as the basis set for our amplitude space dynamics. By subsequently solving for the ker(M) numerically (scipy.linalg.null\_space), we can plot the fixed points in the zero-transients limit of the dynamics. This routine discovers a kernel with dimension 3, the first of the modes is the null vector while the second and third represent superpositions of the basis set modes. Solving numerically for the two kernel modes, we find that the only two states which are steady are the polarized state and the vortex state. This supports the intuition we built from data driven inference in that these two modes represent the foundation of the attracting manifold underlying the behavior of the polarized active elastic sheet animal -- \textit{T. Adhaerens}. 

%\section{The longest length scale modes dominate behavior}
%The self-organization of local energy injection into long-lengthscale coherent dynamics represents an exciting opportunity to study the outcomes of collaborative mechanics in animal behavior. Strikingly, the longest lengthscales are also the modes with the greatest influence on the exploration of the organism. Short lengthscale modes represent a high internal tension or disagreement. These disagreements at the scale of the tissue are mediated by accumulated elastic deformation and the associated restoring force. In sufficiently large organisms the average power density of the states exhibits a power law in the rank (see figure 6). 

\section{Vortex states}
This has been seen in milling dynamics of fish schools\cite{tunstrom_collective_2013}, rotational dynamics of cell clusters \cite{copenhagen_frustration-induced_2018}, and numerous other systems often those in confinement. 
Can we explain these attractors with defect dynamics? 

\subsection{A geometric theory for vortex behavior in small organisms}
Assumptions: transients are short lived and deviations from the manifold in the polarization plane are minimal. 

This gives rise to a simple minimization of elastic energy argument (e.g only the amplitude of the activity disagrees with the vortical state)

Let's begin by assuming that the elastic force balances the difference between the stable velocity and the mean activity. If this is the case, we can write down the contributions via elasticity as:
$$
F_{elastic} = k\delta r \approx \omega (\vec r - \vec r_{COM}) - F_{active}/\gamma
$$

In the limit of linear elasticity, the energy stored in the elasticity takes the form:
$$
E_{elastic} = \frac{F_{elastic}^2}{k}
$$

The next component needed to back of the envelope the energy cost of moving the vortex center away from the center of the organism is the polar equation for a circle which is off-center of coordinates origin.
$$
r' = r_o cos(\theta) + \sqrt{R^2 - r_o^2 sin(\theta)},
$$
where $r_o$ is the distance between the center of the circle and the origin and $R$ is the radius of the circle itself.

Next, we argue that the cilia can adjust its preferred local speed regulated by the height of the tissue. We have shown that in the high curvature limit, this x-z torque has the ability to tune the local activity to match the local velocity. In this application that means there is an annulus defined between $\frac{v_{max}}{\omega}$ and $\frac{v_{min}}{\omega}$ in which the elastic energy stored to maintain this rotating state is approximately zero.

We can break the total elastic energy of this stable vortex into two terms, one which accounts for the tissue that is moving slower than the minimum speed $E_{inner}$ and the tissue that is moving faster than the maximum speed $E_{outer}$.

The functional form governing the change of $E_{inner}$ for the case where $R-r_o > \frac{v_{min}}{\omega}$ take the form of a spatial integral:

$$
E_{inner}(r_o, \omega) = \int_{0}^{2\pi} d\theta \int^{\frac{v_{min}}{\omega}}_0 dr r\frac{( \omega r - v_{min})^2}{k}
$$

This integral can be solved to find (within the limits of $R-r_o > \frac{v_{min}}{\omega}$):
$$
E_{inner}(r_o, \omega) = \frac{\pi}{6 k} \frac{v_{min}^4}{\omega^2},
$$
suggesting that this contribution of the energy (for small $r_o$) simply wants a finite rotation speed to reduce its energy.

The outer energy integral can be written with a similar form and in the regime of small displacements (e.g. $R-r_o > v_{max}/\omega$):
$$
E_{outer}(r_o, \omega) = \int_0^{2\pi} d\theta \int_{v_{max}/\omega}^{r_o cos(\theta) + \sqrt{R^2 - r_o^2 sin(\theta)}} dr r\frac{(\omega r - v_{max})^2}{k}
$$

The simplest perspective takes the viewpoint that the vortex moves relative to a fixed organism position.

Let's solve this in the case where $v_{min} = v_{max}$ so we don't have to break the integral up quite as much for a simple pass.

We will also switch perspectives, instead of the organism moving relative to the vortex, we can compress some of this by solving the vortex moving relative to the organism. To make this simple, it is advantageous to solve in Cartesian coordinates instead of polar. We can exploit the fact that the circular organism is azitmuthally symmetric, which allows us to reduce the dimensionality of the vortex position relative to the center of mass of the organism. 

\begin{figure}
\begin{center}
\includegraphics[width=\textwidth]{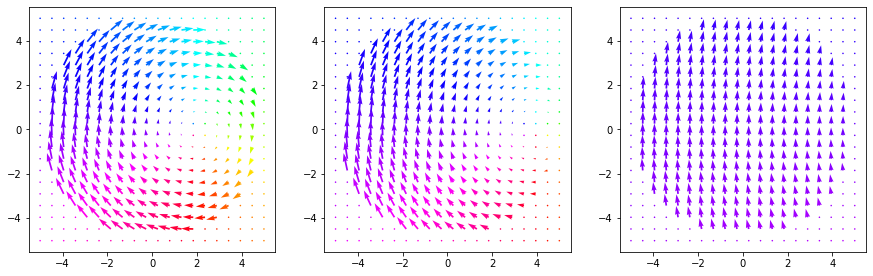} 
\caption{The perspective we adopt is an organism centered at (0,0) of radius R. Inside this region, the displacement field is consistent with a vortex characterized by $\omega$ and $r_o$. Outside of the organism, the contributions are all zero. We may plot three examples of this field for differing values of $r_o$ to give a) $r_o = 0.4R$, b) $r_o = 0.8R$ and c) $r_o = 5R$. For each, the color of the vector signifies the orientation from red-yellow-green-blue-purple-red and the speed of the displacement field is signaled by the vector length.}
\label{fig:vortexfield}
\end{center}
\end{figure}

This integral takes the form of:
$$
E_{elastic} = 2\int_{0}^R dy \int_{-\sqrt{R^2 - y^2}}^{\sqrt{R^2 - y^2}} dx \frac{\left(v- \omega\sqrt{(x-x_o)^2 + y^2}\right)^2}{k}
$$

This accounts for both the disordered regime and the dragged regime in a single integral (while neglecting $v_min$ and $v_max$ satisfied regime). 

This steady state integral of elastic energy gives us a function we can calculate numerically for all positions $r_o$ and $\omega$. 

To make further progress we employ the simple axiom that the system will attempt to minimize stored elastic energy for a fixed $r_o$ position of the vortex relative to the middle of the organism. This principle suggests that, we can write down the $\omega$ that solves for the line minimum of the system for a given $r_o$. 

Mathematically this principle translates into:
$$
\frac{\partial E}{\partial \omega} = 0 = F[\omega_{min}, r_o, R, v]
$$

We can then solve this to find $\omega_{min}(r_o)$. 

Notice that:
$$
\frac{\partial}{\partial \omega}E_{elastic} = 2\int_{0}^R dy \int_{-\sqrt{R^2 - y^2}}^{\sqrt{R^2 - y^2}} dx \frac{\partial}{\partial \omega}\left(\frac{\left(v- \omega\sqrt{(x-r_o)^2 + y^2}\right)^2}{k}\right)
$$

This is very convenient because it allows us to calculate $\omega_{min}$ in the form of the existing integrals.
$$
\frac{\partial}{\partial \omega}E_{elastic} = 0 = \int_{0}^R dy \int_{-\sqrt{R^2 - y^2}}^{\sqrt{R^2 - y^2}} dx \left(v-\omega_{min}\sqrt{(x-x_o)^2 + y^2} \right)\sqrt{(x-r_o)^2 + y^2}
$$

We can then solve for $\omega_{min}$ that satisfies this equation. This gives us:
$$
\omega_{min} = \dfrac{v}{\int_0^R dy \int_{-\sqrt{R^2-y^2}}^{\sqrt{R^2-y^2}}dx \sqrt{ (x-r_o)^2 + y^2}}
$$

\begin{figure}
\begin{center}
\includegraphics[width=0.6\textwidth]{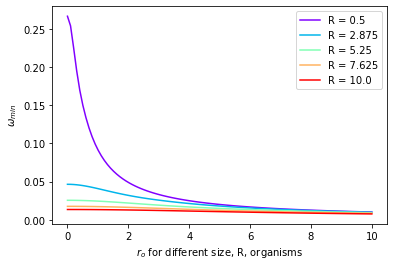} 
\caption{Smaller organisms (red) show a much stronger dependence of rotation speed, $\omega_{min}$ on vortex position, $r_o$ due to their reduced size of the disordered regime.}
\label{fig:vortexomega}
\end{center}
\end{figure}

We find an interesting dependence on the vortex velocity as a function of organism size. Smaller organisms display a much more nonlinear dependence of $\omega_{min}(r_o)$ than larger organisms. This relationship is plotted in figure \ref{fig:vortexomega}.

\begin{figure}
\begin{center}
\includegraphics[width=\textwidth]{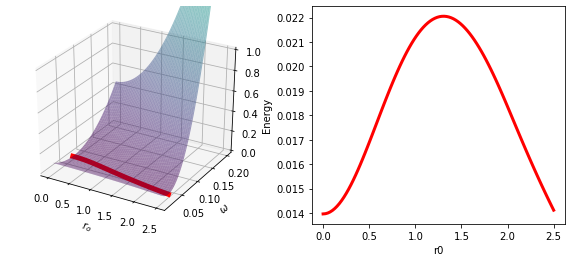} 
\caption{A) The two dimensional energy well is constrained to the red line signaling the predicted (by energy minimization) rotation speed for a given vortex position. B) The resulting reduced dimension of the energy landscape gives a bound vortex which self centers. The escape radius occurs at $\sim 0.65 R$, outside of which, the vortex core is pushed away from the center. This emergent landscape results in a bound vortex.}
\label{fig:vortexenergy}
\end{center}
\end{figure}

The value of this minimization principle is that it allows us to write the minimum elastic energy in terms of the radial distance of the vortex core, by writing $E(r_o, \omega_{min}(r_o))$. Figure \ref{fig:vortexenergy}B shows the resulting energy landscape under the constraint imposed by the energy minimization hypothesis. Two essential features stand out in this energy landscape: 1) the vortex core is self-centering and bound, and 2) the vortex can be ejected once it reaches $r_o \sim 0.65 R$. This result suggest that vortex dynamics of a circular organism can be understood in the zero-transients limit as diffusion on a simple 1D energy landscape. The kinetics of vortex entry and exit can then be approximately reduced to a Kramer's rate process with a complex noise term. 

Our findings that the vortex core is bound in the zero-transient limit fits observations well for small organisms. However, larger organisms (\textit{in silico} and experimental) defy this description. The breakdown of this description with size can be attributed to two reasons: (1) the increased involvement of higher order modes and (2) the increasing timescale of transients with larger and larger organisms. 

The second important implication arises from the interpretation of fluctuations away from the self-centered vortex position. This noise is precisely the dynamics which drive the vortex core away from the center. If this noise grows sufficiently large, this will increase the transition probability of overcoming the bound state energy barrier and sending the vortex off to infinity. From this simple perspective (of a particle diffusing on an energy landscape subject to complex 'noise'), we can characterize the debinding transition as a rate process. It is noteworthy, that in sufficiently high noise environments the +1 defect may delocalize and fail to be simply thought of as a point particle. Below this regime, the noise simply acts to move the vortex core around.

As a first step toward characterizing the noise as a function of system size, we first introduce an important consequence from the inverse energy cascade.

\subsubsection{Defining polarization space}
To characterize the collective dynamics of an ensemble of cells in terms of their longest lived modes (e.g. the only two (meta) stable modes of an active elastic sheet), we define two measurable diagnostic parameters which are related to the underlying order parameters spanning the entire organism: (1) polarization and (2) vortical polarization.

If we consider a 2d space defined by the Polarization of the $\theta$ field and the Vortical Polarization of the $\theta$ field, we can define the bound between the two. This bound tells us the value of each of these fields with zero noise in the system. Any noise will have the effect of pulling the measured state off of this bound.

The $\theta$ field is defined simply as the uniform increase in speed as you move farther away from a vortex center position. Written in terms of the degrees of freedom, this becomes:
$$
\vec\theta(x,y|r_o) = -\omega_{min}y\hat x + \omega_{min}(x-r_o)\hat y
$$

To calculate the polarizations, we only need the orientations of the vectors and thus can drop our calculation of $\omega_{min}$ for the moment.
$$
\hat\theta(x,y|r_o) = \frac{-y}{\sqrt{(x-r_o)^2 + y^2}}\hat x + \frac{x-r_o}{\sqrt{(x-r_o)^2 + y^2}}\hat y
$$

The polarization takes the form of:
$$
P(r_o) = \frac{1}{A} \left| \iint_{A} dA \hat \theta(x,y|r_o) \right|
$$

The vortical polarization takes the form of:
$$
P_{vortical}(r_o) = \frac{1}{A} \left| \iint_{A} dA \hat \theta(x,y|r_o)\times \hat r(x,y) \right|
$$

The first important observation is that these two measures are orthgonal in a sense. When $P_{r_o} = 1$ then $P_{vortical} = 0$. The same is true for $P_{vortical} = 1$ then $P_{r_o} = 0$. However, the seemingly natural assumption that they are directly related as $\sqrt{P_{r_o}^2 + P_{vortical}^2 }= 1$ is not correct. We can instead define the accessible regions of polarization space in terms of the evaluation of both forms of polarization for various values of the vortex center position, $r_o$. 

We begin with the polarization integral:
$$
P(r_o) = \frac{1}{\pi R^2} \sqrt{\left( 2\int_{0}^R dy \int_{-\sqrt{R^2 - y^2}}^{\sqrt{R^2 - y^2}} dx \frac{-y}{\sqrt{(x-r_o)^2 + y^2}} \right)^2 + \left( 2\int_{0}^R dy \int_{-\sqrt{R^2 - y^2}}^{\sqrt{R^2 - y^2}} dx \frac{x-r_o}{\sqrt{(x-r_o)^2 + y^2}} \right)^2}
$$

Next we can figure out the vortical order integral. Recall that:
$$
\vec a \times \vec b = \hat z  (a_xb_y - b_xa_y)
$$
For $\vec a$ and $\vec b$ in a 2D vector field.

This allows us to calculate the integral as

$$
P_{vortical}(r_o) = \frac{2}{\pi R^2} \int_{0}^R dy \int_{-\sqrt{R^2 - y^2}}^{\sqrt{R^2 - y^2}} dx \left( \frac{x(x-r_o) + y^2}{\sqrt{(x-r_o)^2 + y^2}\sqrt{x^2 + y^2}}\right)
$$

\begin{figure}
\begin{center}
\includegraphics[width=0.4\textwidth]{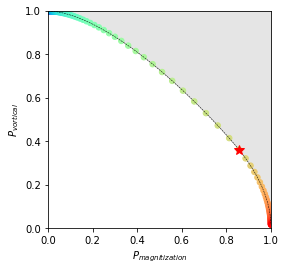} 
\caption{By defining the cellular displacement field at every point of the organism parameterized by $r_o$, we may calculate the corresponding location in this 2D order parameter space defined by the magnetization like polarization, $P_{r_o}$ and a quantity we call the vortical polarization, $P_{vortical}$. Sweeping through $r_o$ as colored from blue to red, we define the manifold in this space which separates a vortex state (with a bound +1 defect) from a spontaneously broken state.}
\label{fig:polspace}
\end{center}
\end{figure}

Solving for these points defines a manifold in Polarization space. When on this manifold, the active elastic sheet is behaving in the simplest possible way, a vortex location (or center of curvature if the vortex is outside the organism) moving around in real space. This gives figure \ref{fig:polspace} illustrating this boundary. Notice that this boundary does not reach the quarter circle defined by $\sqrt{P_{r_o}^2 + P_{vortical}^2 }= 1$. 

The definition of this polarization space manifold opens the discussion of how transients and noise can be considered. It also opens up the characterization of how descriptive we need to be with organisms of increasing size. We can define a simple quantity in this space, the distance from the nearest point on this manifold to the point of our system in time represented in this two dimensional space. We call this $\delta P (t)$. 

$\delta P (t)$ is a measure of how complete the vortex description is of the observed dynamics at any one time. By studying the dynamics of $\delta P (t)$ in addition to the statistics, we find that larger organisms support more complex spatial arrangements and thus have significantly smaller $\langle \delta P (t) \rangle_t$.

\subsection{Toward a delayed field theory which accommodates multiple vorticies}

One way to think about systems where the dynamics of the particle are fast relative to the speed of information transfer in the media in the language of delayed or retarded potentials. 

\subsection{The dynamics of organismal state represented as a +1 defect traveling on a line.}

In the zero-transients limit of a round organism (consistent with intermediate timescales of small organisms), the locomotion kinetics can be represented as a reduced system: a defect core fluctuating on the line orthogonal to the direction of motion. These dynamics can be captured with a simple Langevin equation capturing the dynamics of the point defect on the above defined quasi-static energy landscape. 

\[
\gamma \partial_t r = -\partial_r E(r) + \xi(r)
\]
\begin{figure}[h!]
\begin{centering}
\includegraphics[width = 0.5\textwidth]{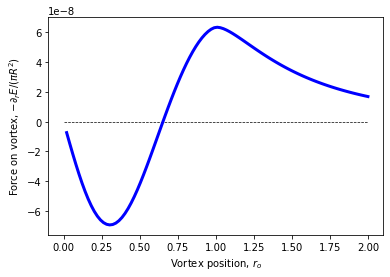}
\caption{The force on the vortex core $-\partial_r E(r)$ is consistent with a bound state characterized by a highly nonlinear storing force up until $0.7R$ after which the vortex is repelled toward $r \rightarrow \infty$.}
\label{fig:vortexforce}
\end{centering}
\end{figure}

The force acting on the vortex as a function of the center of the vortex can be solved through numerical integration to find a functional form which looks quite similar to the derivative of the Lorentzian (with modification to $r<<R$ regime). The resulting curve show in \ref{fig:vortexforce}, intersects zero at $r=0$, $r\approx 0.7 R$ and $r\rightarrow \infty$. 

Under the note-worthy assumption that the noise is approximately white, we can generate microtrajectories of vortex motion by solving this equation using an implementation of Euler-Marayama integration leveraging local solutions of the numerical integration procedure outlined above.

To translate these vortex dynamics into real space locomotion, we develop two simple mappings from vortex position space onto linear and rotational velocity space. The space then becomes $v$ versus $\kappa$. $v = \omega_{min}(r_o)r_o$ and $\kappa = 1/r_o$.

Leveraging the same elastic energy minimization principle used previously, we can also derive $\omega_{min}$. This gives:
$$
\omega_{min} (r_o) = \dfrac{v}{\int_0^R dy \int_{-\sqrt{R^2-y^2}}^{\sqrt{R^2-y^2}}dx \sqrt{ (x-r_o)^2 + y^2}}
$$.

The quality of the fit of this Lorentzian to our full integral form, motivates writing down an approximate form of the dynamics of the vortex.

$$
\partial_t r_{vortex} \approx -\nabla_{r_{vortex}} E(r_{vortex}| R) + \xi(R)
$$

\begin{figure}
    \centering
    \includegraphics[width = 0.5\textwidth]{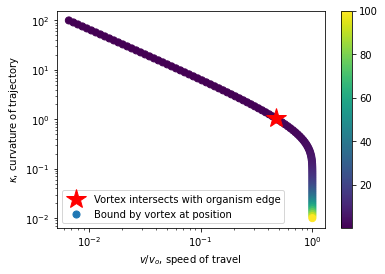}
    \includegraphics[width = 0.4\textwidth]{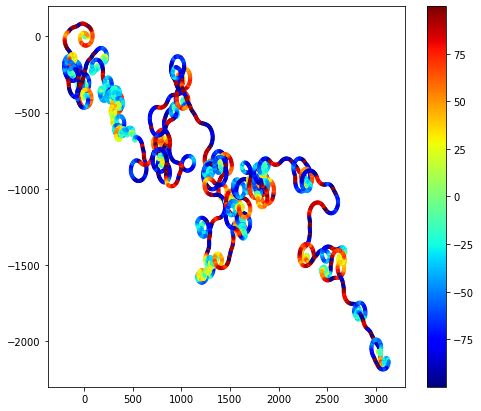}
    \caption{Transforming vortex dynamics onto locomotion. (A) The mapping from vortex core position (relative to organism center) onto center of mass speed of travel and curvature of the trajectory reveals a tight relationship between turning speed and translation velocity which is well approximated by an exponential. Organisms move slowly while turning quickly which is consistent with rotation states in small organisms where they locomote very little. (B) The real-space trajectory of these walks show switching between clockwise and counterclockwise rotation signaled by the red (CW) to blue (CCW) colorbar.  }
    \label{fig:vortextrajectory}
\end{figure}

The resulting dynamics of the organism takes the form:
$$
\partial_t \vec X = v(r_{vortex})cos(\Theta_{COM})\hat x + v(r_{vortex})sin(\Theta_{COM})\hat y
$$
with the angle of displacement of the center of mass takes the form of:
$$
\partial_t \Theta_{COM} = \omega_{min}(r_{vortex})
$$

These direct maps can be differentiated a second time to find:
$$
\partial^2_t x = \frac{\partial v}{\partial r_{vortex}}\frac{\partial r}{\partial t}cos(\Theta) - v(r)sin(\Theta) \frac{\partial \Theta}{\partial t}
$$
$$
\partial^2_t y = \frac{\partial v}{\partial r_{vortex}}\frac{\partial r}{\partial t}sin(\Theta) + v(r)cos(\Theta) \frac{\partial \Theta}{\partial t}
$$
$$
\partial^2_t \Theta = \frac{\partial \omega_{min}}{\partial r}\partial_t r
$$

The velocity curve admits an approximation:
$$
v_{COM}(r | R) \approx v_o (1-e^{\frac{- r}{r_{max}}})
$$
Where $r_{max} \approx \frac{3}{4}R$ is the metastable vortex position.

That also means that $\omega_{min}$, takes the approximate form of:
$$
\omega_{min}(r | R) \approx v_o \frac{(1-e^{\frac{- r}{r_{max}}})}{r}
$$
Where $r_{max} \approx \frac{3}{4}R$ is the metastable vortex position.

The approximation for the energy gradient looks like:
$$
\frac{E}{\pi R^2} \approx \frac{C \Gamma^2}{\left(\frac{r_o}{R}-\frac{r_{max}}{R}\right)^2 + \Gamma^2}
$$

With: $C = 0.0014$,

 $r_{max}/R =  0.67$, 
 
 and $\Gamma =  0.84$.

Plugging these into our system of equations we find:
$$
\partial^2_t x = \frac{\partial v}{\partial r_{vortex}}\frac{\partial r}{\partial t}cos(\Theta) - v(r)sin(\Theta) \frac{\partial \Theta}{\partial t}
$$
$$
\partial^2_t y = \frac{\partial v}{\partial r_{vortex}}\frac{\partial r}{\partial t}sin(\Theta) + v(r)cos(\Theta) \frac{\partial \Theta}{\partial t}
$$
$$
\partial^2_t \Theta = \frac{\partial \omega_{min}}{\partial r}\partial_t r
$$

Let's consider our system as 4 coupled first order differential equations of the form:
$$
\partial_t r \approx \frac{2\pi C R^4\Gamma^2 (r-r_{max}) }{\left((r-r_{max})^2 + R^2\Gamma^2\right)^2} + \xi(R)
$$
$$
\partial_t x \approx v_o\left(1-e^{\frac{- r}{0.75R}}\right)cos(\Theta)
$$
$$
\partial_t y \approx v_o\left(1-e^{\frac{- r}{0.75R}}\right)sin(\Theta)
$$
$$
\partial_t \Theta \approx sgn(r_o) v_o \frac{(1-e^{\frac{- r}{0.75R}})}{r}
$$

A final component to the model suggests that $r= |r_o|$ with the dynamics being the same whether positive or negative. Since $\lim_{r\rightarrow \infty} 1/r = \lim_{r\rightarrow \infty} -1/r$, we also will stitch together $\pm \infty$ using a modulus at a large distance from the origin.

We have demonstrated the nonlinear map from vortex dynamics onto organismal locomotion establishing the critical link between the role of these defects underlying the high variance behavioral modes of this animal.

\section{Controlling defects to harness collective locomotion}
Non-neuromuscular placozoa exhibit effectively low dimensional locomotive dynamics in small organisms where the transients are short compared to the locomotive dynamics. These dynamics can be characterized in by the dynamics of an emergent quasi-particle and a manifold associated with the dynamics of this quasi-particle.

Yet, a critical aspect to these behavioral dynamics is the control of the organismal locomotion, both for the organismal tractability and for our own technological benefit (contributing to our understanding of how to harness the collective modes of active matter to do meaningful work). In this section, we report a tool for directly forcing the defect core through a spatially defined activity gradient. These findings have important ramifications for the controllability of placozoa without direct orientational command. 

\subsection{Steering a vortex by spatial activity control}

Patterning spatial activity has become a powerful tool for shaping the dynamics of active matter systems\cite{zhang_spatiotemporal_2021}. We ask a similar related question, can spatial activity tuning allow placozoa to harness tens of thousands of legs without a coordinating neuromuscular system? We ask this question by studying the quasi-static forcing on the vortex core under an organism wide spatial activity gradient. 

Not only are the vortex within the organism essential for rapid turning dynamics but they illustrate a tool for steering the organism using a spatial activity control (which could be mediated by tissue height dynamics in part 1, chemo-mechanical coupling or even temperature).We can study the dynamics of vortex core manipulation by spatio-temporal activity control by relaxing the radially symmetric assumptions of the previously presented formulation for stable vortex dynamics and by rewriting the immediate forces acting on the vortex with spatially non-uniform activity. 

Leveraging the elastic energy minimization principle (which is consistent with the long time dynamics of the vortex), we can write the instantaneous forcing of the vortex in terms of the energy integral:
\[
E_{elastic} = 2\int_{0}^R dy \int_{-\sqrt{R^2 - y^2}}^{\sqrt{R^2 - y^2}} dx \frac{\left(v(x)- \omega\sqrt{(x-r_o)^2 + y^2}\right)^2}{k}
\]

We can define a very simple functional form of the spatial control of the activity by parameterizing the dynamics to leading order in the x direction:
\[
v(x) = v_o + \delta v*x
\]
Plugging this into our new energy function acts like a constant gradient of activity in the x-direction. The creation of this constant gradient can be thought of in a number of ways ranging from a constant gradient in a activity modulating chemical or the height of the tissue (as specified in part 1).

The instantaneous force acting on the vortex takes the form of:
\[
F_{r_o} = -2\nabla_{r_o} \int_{0}^R dy \int_{-\sqrt{R^2 - y^2}}^{\sqrt{R^2 - y^2}} dx \frac{\left((v_o + \delta v x)- \omega\sqrt{(x-r_o)^2 + y^2}\right)^2}{k}
\]
We can bring our partial derivative inside the integral over the area and evaluate its impact on the integrand:
\[
F_{r_o} = \frac{-2}{k} \int_{0}^R dy \int_{-\sqrt{R^2 - y^2}}^{\sqrt{R^2 - y^2}} dx \frac{2 \omega (x-r_o)\left(v(x)- \omega\sqrt{(x-r_o)^2 + y^2}\right)}{\sqrt{(x-r_o)^2 + y^2}}
\]

Next, we use our proposal that the rotational speed of the body adjusts fast compared to the timescale of vortex motion (which is consistent with the observation in small organisms of a bound vortex state) to extract the value of $\omega$ which minimizes the free energy for a given $r_o$.

We begin finding $\omega_{min}(r_o)$ by evaluating the partial derivative of the energy with respect to the rotational speed.
$$
\frac{\partial}{\partial \omega}E_{elastic} = 2\int_{0}^R dy \int_{-\sqrt{R^2 - y^2}}^{\sqrt{R^2 - y^2}} dx \frac{\partial}{\partial \omega}\left(\frac{\left(v(x)- \omega\sqrt{(x-r_o)^2 + y^2}\right)^2}{k}\right)
$$

This is very convenient because it allows us to calculate $\omega_{min}$ in the form of the existing integrals.
$$
\frac{\partial}{\partial \omega}E_{elastic} = 0 = \int_{0}^R dy \int_{-\sqrt{R^2 - y^2}}^{\sqrt{R^2 - y^2}} dx \left(v(x)-\omega_{min}\sqrt{(x-r_o)^2 + y^2} \right)\sqrt{(x-x_o)^2 + y^2}
$$

We can then solve for $\omega_{min}$ that satisfies this equation. This gives us:
$$
\omega_{min}(r_o) = \dfrac{\int_0^R dy \int_{-\sqrt{R^2-y^2}}^{\sqrt{R^2-y^2}}dx \left(v(x)\sqrt{(x-r_o)^2 + y^2}\right)}{\int_0^R dy \int_{-\sqrt{R^2-y^2}}^{\sqrt{R^2-y^2}}dx \left[ (x-r_o)^2 + y^2\right]}
$$

This suggests that the spatial modulation of the activity modifies both the rotational velocity of the vortex. 

This gives us in the instantaneous force an activity gradient applies to a vortex core in the quasi-static limit:
\[
F_{r_o} = \frac{-2}{k} \int_{0}^R dy \int_{-\sqrt{R^2 - y^2}}^{\sqrt{R^2 - y^2}} dx \frac{2 \omega_{min}(r_o, \nabla_x v) (x-r_o)\left((v_o + \nabla_x v\cdot x)- \omega_{min}(r_o, \nabla_x v)\sqrt{(x-r_o)^2 + y^2}\right)}{\sqrt{(x-r_o)^2 + y^2}}
\]
and
\[
\omega_{min}(r_o, \nabla_x v) = \dfrac{\int_0^R dy \int_{-\sqrt{R^2-y^2}}^{\sqrt{R^2-y^2}}dx \left((v_o + \nabla_x v \cdot x)\sqrt{(x-r_o)^2 + y^2}\right)}{\int_0^R dy \int_{-\sqrt{R^2-y^2}}^{\sqrt{R^2-y^2}}dx \left[ (x-r_o)^2 + y^2\right]}
\]

We can solve for these relationships numerically to explore the role of spatial activity gradients on steering the organism by way of the effective force on the vortex core.

\begin{figure}[h!]
\centering{
A)\includegraphics[width =0.34\textwidth]{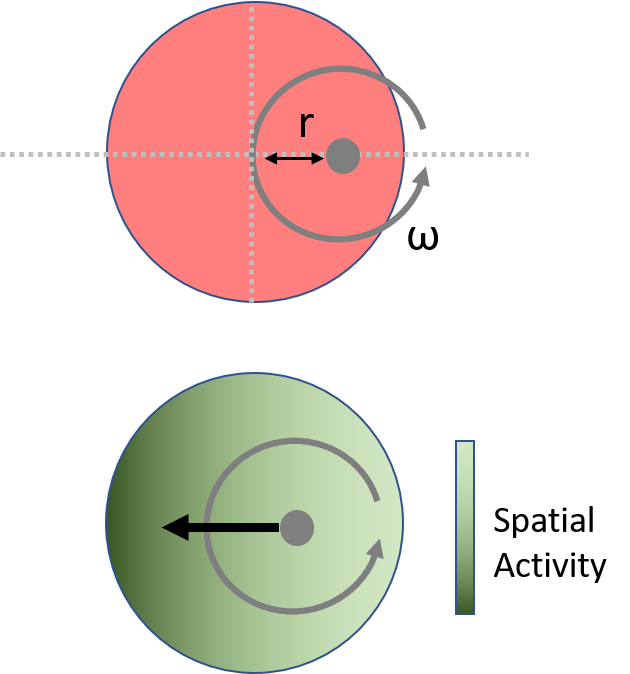}
B) \includegraphics[width =0.58\textwidth]{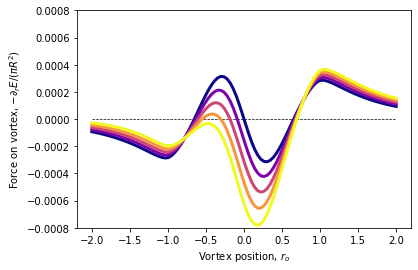}
\caption{\textbf{Spatial activity control guides vortex core motion.} A) The vortex core motion can be considered an emergent quasi-particle in quasi-steady state characterized by a position and a rotational speed both of which can be calculated using our tissue elasticity energy minimization. The addition of a spatially encoded activity can be written to lowest order as a uniform gradient $v(x) = v_o + \nabla_x v \cdot x$, where the cells on the left are mapping a higher chemical signature to a lower preferred activity than the cells on the left. B) We show that for small spatially encoded activity gradients $\nabla_x v = \{0 =\text{blue}, ... , 7\% v_o = \text{yellow}\}$ the effective force acting on the vortex core as a function of position. The size of the stable region decreases rapidly with increasing activity gradient and destabilizes completely when the activity gradient exceeds a critical, size-dependent value.}
\label{fig:gradV_forceOnCore}
}
\end{figure}

Activity gradients are a powerful tool for the organism to manipulate the direction of travel through the effective force placed on the defects. In figure \ref{fig:gradV_forceOnCore}B), we show that the effective force on the vortex core opposes the gradient and above a critical threshold can destabilize the bound state at the center of the organism. This means that the defect core will reliably travel down the activity gradient. 

The direct mapping between organismal locomotion and defect dynamics has important ramifications for the effective interaction of the organism's direction of travel with the establishment of a spatial activity gradient. Due to the relationship between defect direction and radius of organismal curvature established in \ref{fig:vortextrajectory}, this means that a tribotaxing organism will have a strong tendency to move perpendicular to the spatial gradient in activity. 

The control of defects in the collective mechanics of placozoan locomotion can serve as a powerful tool for both us and the organism to steer the tissue on the basis of only locally evaluated fields of chemomechanical or mechanical coupling.

\subsection{Defect stability and dynamics can be controlled by organism shape}

Placozoa are dynamic shape shifters \cite{prakash_motility-induced_2021}. In this section, we illustrate how leading order changes to shape (under conserved number of cells) change the steady-state dynamics of the mobile defects which propagate on to the locomotive behavior of the entire organism. 

\begin{figure}[h!]
\centering{
A)\includegraphics[width =0.28\textwidth]{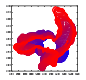}
B)\includegraphics[width =0.24\textwidth]{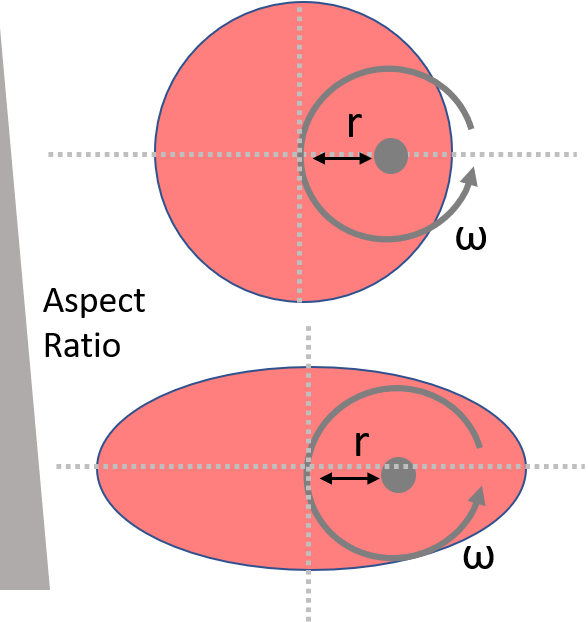}\\
C) \includegraphics[width =0.6\textwidth]{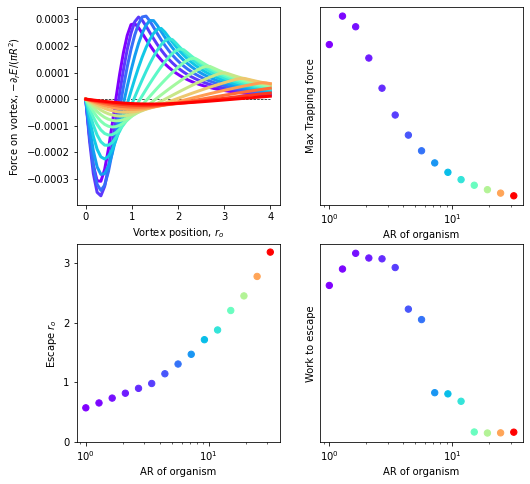}
\caption{\textbf{Shape alters the force landscape on mobile defects.} A) Placozoa dynamically change shape over 10's of minute timescales in response to ciliary forcing and contractility. B) We study the emergent force landscape acting on defects as a function of the organism's eccentric shape, summarized by the chosen aspect ratio for conserved area. C) Aspect ratios of 1 (purple) recreate the existing force landscape while larger aspect ratios flatten and extend the force landscape. In intermediate aspect ratios the maximum trapping force increases before decreasing correlating with work for the vortex to escape increases over an intermediate region of eccentricity.}
\label{fig:AR_forceOnCore}
}
\end{figure}

We expand our steady-state defect model to include non-circular organisms by including the leading order contributions to eccentricity. To control for organism size, we use a simple definition for the organism shape of the area preserving form:
\[
ARx^2 + \frac{y^2}{AR} = R^2
\]
which controls the region of integration for the defect-parameterized, steady-state displacement field.

In figure \ref{fig:AR_forceOnCore}, we show the alteration of the force landscape with respect to change in position of the defect core and extract simple summary metrics for the evolution with changing aspect ratio. At large aspect ratio ($AR\sim10$) the work needed for the defect to escape approaches zero. However, we also report that at small aspect ratios $AR\sim 2$, the defect-debinding energy grows before it shrinks. Intuitively, this growth for small aspect ratio can be understood as a property of the position weighted sum of the energy. Since the total amount of tissue traveling at high speeds is larger for a $AR=2$ tissue than the circular tissue the stabilizing force for moving away from the center of the tissue increases.

The control of defects by way of organismal shape control suggests another possible route by which the locomotive characteristics of the organism can be modified by biologically controllable knobs. High aspect ratio organisms destabilize defects cores and incentivize time spent in other locomotive modes perhaps less prone to rotation states where the organism doesn't walk away in a ballistic-diffusive trajectory.

\section{Discussion}
Our objective in this work has been threefold: (1) identify the dominant spatio-temporal modes underlying the collective locomotion of a polarized active elastic sheet (motivated by \textit{T. Adhaerens}), (2) connect the physical mechanism that shapes these emergent dynamics and (3) derive a deeper understanding of the resulting manifold. Along the way, we have presented a few more general lessons including the importance of high variance modes, one mechanism for deriving emergent manifolds from emergent many-body systems and tools for controlling the defects with direct consequences on locomotive behavior of the organism.

\subsection{High variance modes signal locomotion while low variance modes harness simple decision making}

We have shown that through simple unsupervised techniques that the effective dynamics of a polarized active elastic sheet stimulated by pink activity noise can be captured with the study of relatively few modes. This supports a natural separation between the high variance and low variance modes of the decomposition. Conceptually, the high variance modes carry the bulk of the locomotive information while the low variance modes can influence the locomotive dynamics only through modification of the high variance modes. 

The role of the high variance modes is reminiscent of the neural dynamics in the primate motorcortex where grasping events are punctuated by striking low-dimensional projections of the sampled neurons activity space\cite{vyas_computation_2020}. Here the low variance modes are largely discarded as noise. 

In our work, we find that a very different physical system leverages the coupling between the high variance and the low variance modes to evoke higher order behavior and we suggest that this motivates deeper study of the residual modes left behind after subjecting off the high variance contributions. We hope this this approach might lend deeper insights into the dynamics near the decision boundary in plaoczoa and perhaps beyond. This direct could hold important insights into the emergent statistics of the 'noise' terms in our low dimensional dynamics.

\subsection{Connecting the emergent manifolds to a mechanism}
We subsequently looked to see if the documented dynamics of the modes of the active-elastic network could shed light on the emergence of these effectively low dimensional dynamics. We found that this energy cascade approaches two states in the zero-transients limit: the polarized state and the vortex state. This result fits nicely into the existing literature which has previously identified rotation, flocking and disorder as the dominant modes of a collective fish \cite{tunstrom_collective_2013, attanasi_information_2014}, birds \cite{attanasi_information_2014}, insects\cite{van_der_vaart_environmental_2020} and cellular systems\cite{copenhagen_frustration-induced_2018}. Our contribution to the understanding of an active elastic system helps to refine our understanding of what shapes these manifolds and how microscale biophysics can be mapped up to effective organismal behavior through a process of explicit mode coupling mediated through activity. 

Our hope is that the paradigm presented in this simple animal may contribute in a small way to the reinforce the community's conversation surrounding not just what the low dimensional behavior manifolds look like but how and perhaps why they are shaped\cite{berman_predictability_2016, ahamed_capturing_2021, stephens_dimensionality_2008, daniels_automated_2019}.

\subsection{Controlling defects}
The active matter community has long used defects as a means for reducing the dimensionality of the system under study \cite{marchetti_hydrodynamics_2013, saw_topological_2017, shankar_topological_2020}. More recent pushes has linked these defect dynamics to the emergent behavior of the underlying system \cite{peng_command_2016, saw_topological_2017} and even manipulating emergent particles in programmable arenas\cite{zhang_spatiotemporal_2021, gong_engineering_2020}. We look forward to the continued cross-pollination between the emergent low dimensional manifolds sketched out by defect dynamics and the resulting organismal behavior. Perhaps there are more circumstances in the tree of life where the two can profitably meet in the middle to deeper our understanding of the effective dynamics in organism-scale decision making.

\section{Conclusion}

With a concerted choice in model organism, we have been able to map population dynamics of tens of thousands of dynamically rich cells (e.g. each equipped with a walking cilium capable of bistable switching \cite{bullpart1} and a continuous internal orientation state\cite{bullpart2}) onto low order models which capture the bulk of the locomotive behavior of small placozoa. We have shown how these effective low dimensional models arise from a physical self organization of the locomotive activity through an active-elastic inverse energy cascade. The pumping of energy from the shortest lengthscale into collective modes is a hallmark of active matter\cite{alert_active_2021} and the population dynamics of simple neuronal systems\cite{pouget_information_2000}. Merging these two ideas may be an interesting way forward toward an alternative understanding of how the local rules shape the emergent manifolds. 

A natural followup conjecture is that similar Cascades in mode amplitude space may not simply govern the long-time scale steady state of the dynamics of this distributed sensory-actuator behavioral system, but that the transients of a sub-population of the low-variance modes may evoke rich dynamics sufficient to support reservoir/embodied computation \cite{BullPerspectiveInPrep, gilpin_multiscale_2020}.

The physics of behavior community has made tremendous progress on tools for discovering low-dimensional descriptions. This abundance of data from diverse systems is poised to generate deep insight into not just what separates these systems but also want unifies elements of animal behavior ranging across the diverse spectrum. This discussion can only be furthered by a complementary perspective in a simpler system such as the non-neuromuscular placozoa. And yet, there is an opportunity to learn from how this simple system uses a biomechanical design language to control its rapid behavior transitions without a central controller. This paradigm of distributed senses and actuation promises a complementary view of the dynamics of populations of cells in service of the whole animal.

Our work also illustrates the limits of this approach in that it reveals a proliferation of relevant active-elastic modes which scales with organism size. These size dependent dynamics reveal both an enrichment of the locomotive complexity as the organism grows, but also provides an experimental measurement technique for capturing the deviation of the observed dynamics from the effective low-dimensional manifold proposed to capture these dynamics. 

When viewed from the perspective of trapped vortex states, ascending cascades of active-elastic modes and spontaneously broken symmetries, the dynamics the mechanics of a living fossil begins to inform some of the governing principles of collective dynamics of large ensembles of cells. These studies provide a concrete and easy to understand terrain for probing how to harness these collective mechanics in service of an animal's survival in a highly dynamic environment. 

We expect that this work will be of interest to engineers building with active matter and those seeking general principles governing the physics of animal behavior.

\section{Acknowledgements}
We thank all members of the PrakashLab for scientific discussions and comments. In particular, we thank Vivek Prakash, Shahaf Armon, Pranav Vyas, Grace Zhong, Hazel Soto-Montoya and Laurel Kroo for discussions. M.S.B. was supported by the National Science Foundation Graduate Research Fellowship (DGE-1147470) and the Stanford University BioX Fellows Program. This work was supported by HHMI Faculty Fellows Award (M.P), BioHub Investigator Fellowship (M.P), Pew Fellowship (M.P), Schmidt Futures Fellowship, NSF Career Award (M.P), NSF CCC (DBI-1548297) and Moore Foundation. 

%Clarice Aiello for support in optical configuration and software of tracking microscope. Rudro Iyer-Biswas,  Fabien Pease, Jan Skotheim, Thomas Perkins for fruitful scientific discussions and scientific mentorship.  We thank Alex Williams for introducing us to methods and software for pattern recognition in spatio-temporal fields. Toly [], Brandon [flocking, image registration, coupled oscillations, excitable dynamics], Guillermina [ciliary and tissue physics], Haripriya, Scott [ciliary signaling], Grace [], Micheala [for insightful discussions of adhesion physics], Anton [frustration and numerical methods], William [synchronization, computer vision and XY models], Lucas [synchronization], Shahaf [coupled oscillators, tissue physics and experimental protocols], Arnold [active matter theory and hydrodynamics near interfaces], TingTing [nematic active matter and controlling distributed actuation], Felix [critical questioning and behavioral ethnography] as well as all members of the Prakash lab past and present [fix order]. Funding to MSB, NSF-GRFP, BioX fellow. M.P. [add in]

\bibliography{mybib2.bib}

\end{document}